\newcommand{\beq}{\begin{equation}}
\newcommand{\eeq}{\end{equation}}
\newcommand{\dd}{{\rm d}}
\newcommand{\fig}[1]{Fig.\,\ref{#1}}
\newcommand{\eqn}[1]{Eq.\,(\ref{#1})}
\newcommand{\sect}[1]{Sect.\,\ref{#1}}
\newcommand{\tab}[1]{Table\,\ref{#1}}
\newcommand{\appx}[1]{Appendix\,\ref{#1}}

\documentclass{aa}  

\usepackage{xcolor}
\usepackage{graphicx}
\usepackage{txfonts}
\usepackage{dsfont}
\usepackage{mhchem}
\usepackage{environ}
\usepackage{array}
\usepackage{hyperref}
\hypersetup{
    colorlinks,
    citecolor=blue,
    filecolor=blue,
    linkcolor=blue,
    urlcolor=blue
}
\usepackage{datetime}

\usepackage{soul}

\usepackage{orcidlink}

\definecolor{orange}{rgb}{1,0.5,0}
\definecolor{sred}{rgb}{.5,0,0}

\newcommand{\reply}[1]{{#1}}

\newcommand{\theneutral}{$p$-\ce{NH2D}}
\newcommand{\theion}{\ce{N2D+}}

\begin{document}

   \title{A differentiable and optimizable 3D model for interpretation of observed spectral data cubes}


   \author{T.~Grassi
          \inst{1,2}\fnmsep\thanks{Corresponding author,  \email{tgrassi@mpe.mpg.de}}\orcidlink{0000-0002-3019-1077}
          \and
          J.~E.~Pineda\inst{1}\orcidlink{0000-0002-3972-1978}
          \and
          S.~Spezzano\inst{1}\orcidlink{0000-0002-6787-5245}          
          \and
          D.~Arzoumanian\inst{3,4}
          \and
          F.~Lique\inst{5}
          \and
          Y.~Misugi\inst{6}
          \and
          E.~Redaelli\inst{7}\orcidlink{0000-0002-0528-8125}
          \and
          S.~S.~Jensen\inst{1}\orcidlink{0000-0003-2480-0742}
          \and
          P.~Caselli\inst{1,2}\orcidlink{0000-0003-1481-7911}
          }

   \institute{Max-Planck-Institut f\"ur Extraterrestrische Physik, Giessenbachstra{\ss}e 1, 85748 Garching, Germany 
         \and
             ORIGINS Excellens Cluster, Boltzmannstra{\ss}e 2, 85748 Garching, Germany
          \and
          Institute for Advanced Study, Kyushu University, Japan
          \and
            Department of Earth and Planetary Sciences, Faculty of Science, Kyushu University, Nishi-ku, Fukuoka 819-0395, Japan
            \and
              Univ Rennes, CNRS, IPR (Institut de Physique de Rennes) - UMR 6251, F-35000 Rennes, France
          \and
          Faculty of Science and Engineering, Kyushu Sangyo University, 2-3-1 Matsukadai, Fukuoka 813-8503, Japan
          \and
          European Southern Observatory, Karl-Schwarzschild-Stra{\ss}e 2, 85748 Garching, Germany
             }

   \date{Received -; accepted -}

 
  \abstract
   {}
   {Molecular spectral cubes of prestellar cores encode the information on the physical and chemical properties of these objects along the line of sight. To retrieve this information, we need an interpretable model that reproduces the observed spectra.}
   {We designed a differentiable 3D geometrical model that produces synthetic observations from the parameterized density and velocity fields, and that can be efficiently optimized to reproduce the real data cubes. The model has been applied to \theneutral{} and \theion{} spectral cubes in the prestellar core L1544.}
   {The optimized model suggests that to reproduce the observed velocity difference between \theneutral{} and \theion{} in L1544, an asymmetric structure in density and velocity is necessary.}
   {}


   \maketitle
%
\section{Introduction}\label{sect:introduction}
Observations of prestellar cores \citep{Caselli2002} provide increasingly more detailed spectral-line and continuum data that encode the detailed structure of these objects, including their density \citep{Lin2022}, temperature \citep{Crapsi2007,Harju2017,Pineda2022}, kinematics \citep{Caselli2002,Keto2008,Punanova2018,Redaelli2022b}, geometry \citep{Tritsis2016}, and chemical budget \citep{Bergin2005,Spezzano2017}.

To connect the observations of star-forming regions and more generaly that of the interstellar medium to their actual physical properties, several models have been proposed, including, for example, the use of emission lines to determine various physical parameters \citep{Kaufman1999,Pety2017}, a range of chemical tracers to obtain the cosmic-ray ionization rate \citep{Caselli1998,Padovani2024,Redaelli2024}, chemo-dynamical models to determine the kinematic properties \citep{Keto2015,Sipila2018,Sipila2022}, and a series of chemical post-processing of hydrodynamical simulations to explore the chemical properties in general \citep{FerradaChamorro2021,Jensen2023,Panessa2023,Priestley2023,NavarroAlmaida2024,Narayan2025}. 

Recently, these classical approaches have been broadened to machine learning methods, including deep reinforcement learning \citep{Qiu2025}, interpretable machine learning methods \citep{Heyl2023,AsensioRamos2024,Diop2024,Grassi2025,Vermarien2025a,Vermarien2025b}, deep neural networks \citep{Behrens2024,Kessler2025,Morisset2025}, Bayesian methods \citep{DeCeuster2023,Heyl2023b,Lin2025,Palaud2025}, and autoencoders \citep{Portillo2020,ShafaatMahmud2025}.

In addition to these relatively classic methods, the recent development of differentiable models has emerged as a natural evolution of machine learning computational architectures \citep{GunesBaydin2015, Innes2019,Cranmer2020}. A differentiable model represents a mapping from inputs to outputs with a smooth dependence, i.e., a small change in the parameters induces limited and predictable changes in the model output. This property enables the computation of gradients to determine the variations of the outputs with respect to each parameter \reply{and, in addition, it is employed to accelerate Bayesian inference workflows, for example by enabling gradient-based samplers such as Hamiltonian Monte Carlo \citep{Phan2019,Campagne2023,Cabezas2024}}. When target data is employed, these gradients suggest how each parameter should be adjusted to minimize the difference between predictions and observations. 

A differentiable model can therefore be included directly in the optimization pipeline, rather than using its emulators or some approximated representation, in the so-called solver-in-the-loop fashion \citep{Um2020}. However, most of the computational models employed in astrophysics are not differentiable, with some notable exceptions, like \texttt{jf1uids}, a one-dimensional differentiable fluid solver \citep{Storcks2024}, \texttt{diffhydro}, a fully differentiable hydrodynamical code with multiphysics \citep{Horowitz2025a,Horowitz2025b}, \texttt{Ray-trax}, a 3D ray tracing code designed to be integrated into differentiable hydrodynamical solvers \citep{Branca2025}, \texttt{RadJAX}, a line radiative transfer solver \texttt{} \citep{Levis2025}, and \texttt{Carbox} a differentiable chemical network solver \citep{Vermarien2025c}.

\reply{Another notable exception, directly related to the model we propose here, is \texttt{RUBIX}, an end-to-end differentiable code designed to forward model integral field unit spectrograph cubes of galaxies from cosmological hydrodynamical simulations \citep{Cakir2024,Schaible2025}. The main difference is that while \texttt{RUBIX} is based on simulation outputs, our model directly parameterizes the background model within a differentiable framework, enabling exploration of several geometrical configurations on the fly. Although this is an advantage during our optimization process, the drawback is that our approach ignores the fully consistent physical constraints that a hydrodynamical simulation encodes by construction.}

Most of the above codes rely on JAX \citep{Jax2018}, a Python library designed for high-performance calculations, specifically for GPU-accelerated array computation. A key characteristic of JAX is that it can automatically differentiate native Python and NumPy functions, supporting reverse-mode differentiation (i.e., backpropagation). We leverage these two characteristics in our work, where a differentiable parameter-controlled 3D geometrical model is employed to generate synthetic observations. The differentiable fashion allows for optimizing the model against observed data.

In \sect{sect:model}, we present the model and its capability of producing synthetic observations, while \sect{sect:fitting} describes the optimization loop employed on the input parameters. In \sect{sect:results}, we compare the obtained data with the actual observations, and we propose an interpretation based on the optimized model. \sect{sect:conclusions} illustrates the conclusions.

\section{Numerical model and synthetic observations}\label{sect:model}
In this paper, we employ a differentiable pipeline that relies on a 3D analytical physical model controlled by a set of parameters and generates a corresponding synthetic spectral cube\footnote{Example applications of the code can be found at \url{https://github.com/tgrassi/ppv_jax_public} and described in \appx{sect:test}.}. In this specific context, for the sake of clarity, we will refer to the case of two spectral cubes (also known as position-position-velocity cubes, or abbreviated as PPV cubes) observed toward L1544, one for a neutral molecule (\theneutral, $J_{Ka\,Kc}$=$1_{11}$-$1_{01}$) and another for an ion (\theion, $J$=2-1). Further details about the observed data are reported in \sect{sect:results}.

\begin{figure*}
\centering
    \includegraphics[width=0.9\textwidth]{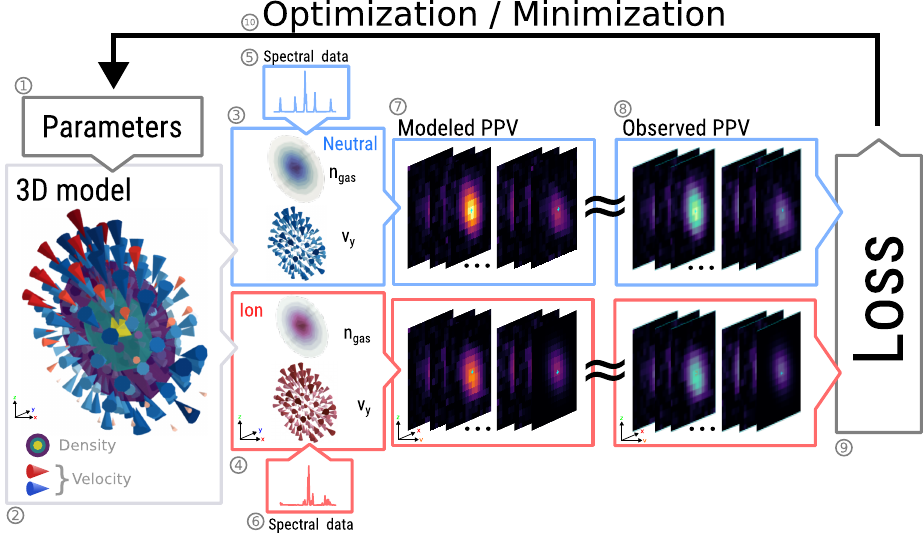}
    \caption{Sketch of the methods employed. Given a set of parameters (1), we generate a 3D model (2) that contains the velocity field along the LOS and the number density information for \theneutral{} (3) and \theion{} (4). With this 3D data information, we can utilize their known spectral emission features (5 and 6) to compute the emission and absorption in each velocity channel, thereby generating a set of modeled PPV spectral cubes for each species (7). We compare the generated PPV with the observed one (8), computing a loss function (9) that will be used to modify the parameters in order to minimize the loss (10). The cycle is repeated until the loss is minimized. }
        \label{fig:sketch}
\end{figure*}

The core concept is to compare synthetic observations with real ones and optimize the model parameters to minimize the difference between them. The method is sketched in \fig{fig:sketch}, where a set of parameters (1) is employed to construct a 3D analytical model (2) to determine the density and velocity distribution of a neutral molecule (3) and an ion (4). Using the molecular spectroscopic data, radiative transfer is performed (5, 6) to produce synthetic spectral cubes (7). These are compared to their observed counterparts (8) to produce a loss function (9) that is minimized in order to optimize the initial parameters (10). The loop is repeated until the loss is below a chosen threshold.

To this aim, we use JAX\footnote{Version 0.6.2, \url{https://github.com/jax-ml/jax}} as our preferred framework due to its efficiency and intrinsic differentiability. This allows to efficiently generate synthetic observations from a given set of model parameters, and for easy integration of the pipeline into \textsc{Optax}\footnote{Version 0.2.6, \url{https://github.com/google-deepmind/optax}} \citep{Optax2020}, an optimization framework designed to be seamlessly integrable with JAX.

The analytical model for the physical properties is a 3D grid of size ($N_x$, $N_y$, $N_z$), where at each point we compute the three velocity components $(\varv_x, \varv_y, \varv_z)$ and the density ($n$) of the two molecules. The second dimension ($y$ in the observer coordinate system) is assumed to be the line-of-sight (LOS) of the observer, and we usually assume that $N_y \gg N_x=N_z$, i.e., higher spatial resolution along the $y$ component, since the gas properties will be integrated along this LOS, and the other coordinates are linked to the spatial resolution of the observations. While the observer coordinate system is fixed, ${\bf x}\equiv(x, y, z)$, the object coordinate system, ${\bf x}_{\rm o}\equiv(x_{\rm o}, y_{\rm o}, z_{\rm o})$, is allowed to rotate\footnote{The rotation is to be considered with respect to the observer frame of reference.} following its intrinsic rotation system (also known as yaw, pitch, and roll angles indicated here as $\vartheta_{\rm y}$, $\vartheta_{\rm p}$, and $\vartheta_{\rm r}$), and to rigidly translate (parameters $x_{\rm c}$ and $z_{\rm c}$), but only in the $x$ and $z$ coordinates of the observer system (i.e., it cannot move along the LOS coordinate $y$, but only on the plane of the sky). Finally, the domain is $x=y=z\in[-1,1]$ in code units\footnote{Code length units $L$ can be translated to physical units, considering the observer's distance from the object $d$ and the angular size of the spectral cube $\alpha$, as $L=d\tan(\alpha/2)$.}.

\subsection{Velocity field}
To determine the actual values of the velocity components, $\mathbf{v}\equiv(\varv_x,\varv_y,\varv_z)$, we first assume a versor\footnote{A vector with norm one.} field $\mathbf{u}=M\mathbf{x}$ determined by the 9 elements of a matrix $M$, where $\mathbf{x}$ are the coordinates at a given position. For example, a symmetric free-fall collapse will be represented by $M=-I$, where $I$ is the identity matrix. $M$ is a global matrix, i.e., it does not depend on the position $\mathbf{x}$.  The versor $\mathbf{u}$ indicates the direction of the velocity field, while the actual velocity magnitude is controlled by a radial function
\beq\label{eqn:velocity}
 \varv(r) = \frac{\varv_0}{\sigma_\varv\sqrt{2\pi}} \exp\left[-\frac{(r-r_\varv)^2}{2\sigma_\varv^2}\right] \cdot \left[1+a_\varv\sin(\varphi+\varphi_0)\right]\,,
\eeq
where $r=\sqrt{x_{\rm o}^2+y_{\rm o}^2+z_{\rm o}^2}$ is the radius in the object frame, $\varv_0$, $r_\varv$, and $\sigma_\varv$ are respectively the amplitude, the position, and the width of a Gaussian function, while $a_\varv$ and $\varphi_0$ are respectively the amplitude and the phase of a sinusoidal perturbation along the $z_{\rm o}$ symmetry axis in the object framework, where $\varphi=\tan^{-1}(y_{\rm o}/x_{\rm o})$ is the azimuthal angle\footnote{Implemented as \texttt{arctan2}.}. This perturbation is introduced to model large-scale asymmetries, and therefore, we limit the sine cycles to a single cycle. Finally, we add a velocity $\varv_{\rm bulk}$ along the LOS component ($y$), that is the collective velocity of the object with respect to the observer. The actual velocity vector is then $\mathbf{\varv}(r) = (0, \varv_{\rm bulk}, 0) + \varv(r)\,\mathbf{u}$.

It is worth noting that since there are no physical constraints, the only ``observed'' velocity component is the $y$ component, i.e., along the observer LOS. The only effect of the other components is in the normalization of ${\bf u}$, and therefore non-physical configurations in the $x$ and $z$ velocity components might be found, but, in principle, it is possible to constrain the velocity field, for example, to avoid outflows or other unrealistic solutions, as shown later.

\subsection{Density distribution}
The number density distribution of a given molecule is modelled analogously by using a radius-dependent function
\beq\label{eqn:density}
 n(r^*) = \frac{n_0}{\sigma_n\sqrt{2\pi}} \exp\left[-\frac{(r^*-r_n)^2}{2\sigma_n^2}\right]\,,
\eeq
where $n_0$, $r_n$, and $\sigma_n$ are respectively the amplitude, the position, and the width of a Gaussian function, while $r^*=\sqrt{x_{\rm o}^2+y_{\rm o}^2+\delta^2z_{\rm o}^2}$, i.e., the radius of an ellipsoid controlled by a parameter $\delta$ that determine the sphericity ($\delta=1$), the oblateness ($\delta>1$), or the prolateness ($\delta<1$) of the gas density distribution. The radius is calculated using object coordinates. It is important to note that here, the density does not necessarily refer to the actual number density of the given chemical species, but rather to the abundance of the molecule emitting it in a specific transition, i.e., an effective perfect emitter. In other words, in this model, a non-emitting molecule for a specific transition will have zero density, and for the same reason, a different transition of the same molecule will be mapped by a different number density distribution. In theory, it would be possible to define an actual chemical number density using a differentiable chemical solver (e.g., \texttt{Carbox} \citealt{Vermarien2025c}) or specifically designed lookup tables, but this is beyond the scope of this paper. In practice, this means that all the hyperfine transitions have the same excitation temperature, which is a standard approach \citep{Harju2024,Pineda2026}.

\subsection{Molecular lines emission and absorption}\label{sect:emission}
With these parameterizations, for each molecule, we obtain four three-dimensional grids (three for each velocity component and one for the density). However, for the synthetic observations, we are only interested in the $y$ component of the velocity ($\varv_y$). This represents the velocity of each cell with respect to the observer's LOS. Each grid element is assumed as a locally uniform volume of emitting gas with density $n$, composed of molecules that exhibit multiple known emission features at rest. To this aim, we obtain their relative velocity shifts ($\Delta \varv_i$) and their relative intensities ($I_i$) from \textsc{PySpecKit}\footnote{See \texttt{spectrum/models/nh2d.py} (commit \texttt{50348a1}) and \texttt{spectrum/models/n2dp.py} (commit \texttt{41a17e7}), in the GitHub repository at \url{https://github.com/pyspeckit/pyspeckit}.} \citep{Ginsburg2011,Ginsburg2022} for both \theneutral{} \citep{Daniel2016,Melosso2021} and \theion{} \citep{Dore2004,Pagani2009}. \reply{It is worth noticing that we use the data from the \textsc{PySpecKit} database, because they are already in the convenient form of Python dictionaries and they are curated by the code community. In other words, \textsc{PySpecKit} is not included in the differentiable pipeline. These data} allow us to construct the actual emission profile as the sum of Gaussian functions centered at $\varv_y+\Delta \varv_i$ with width $\sigma_{\rm t}$ and amplitude $I_i$.
Scaling the total emission with the local tracer density $n$, we obtain the amount of the emitting gas at a given grid point and within a given velocity bin (i.e., brighter lines have a larger weight in the final emitting material density). These functions are computed on top of a predefined velocity grid that represents the velocity channel binning. Using an $N_\varv$-channel spectrum, we obtain a new grid of size ($N_x$, $N_y$, $N_z$, $N_\varv$), which represents the emitting gas intensity in each spectral channel and spatial grid point.

However, we assume that the gas also absorbs the emitted radiation, depending on its velocity and spectral features.
The final emission intensity in a velocity channel centered at $\varv$ that reaches the observer from any point along the LOS is then
\beq\label{eqn:emission}
 E_i(\varv) = w_i(\varv)\exp\left(-\sum_{j=0}^i w_j(\varv)\, \sigma_{\rm abs} \Delta y\right)\,, 
\eeq
where the absorption is computed as a sum from the observer position ($j=0$) to the given $i$th emission point, where $\sigma_{\rm abs}$ is a free parameter, and $\Delta y=2\,N_y^{-1}$ in code spatial units is the spatial step along the LOS\footnote{Since the absorption in \eqn{eqn:emission} is the argument of an exponential function, to reduce the sensitivity to this specific parameter and improve the optimization in the code we use $\log_{10}(\sigma_{\rm abs})$.}. The term $w$ represents the effective density of the molecule at a specific velocity channel at a given position. We note that with this approximation, the ``emitting density'' ($w_i$) and the ``absorbing density'' ($w_j$) are the same, scaled by a constant factor $\sigma_{\rm abs}\Delta y$, while in principle, we should compute the level population of the given molecule and consider its actual properties, resulting, as an effective variable scaling along the LOS. We tested this scenario by employing an additional parameterization that models the ``absorbing density'' with the equivalent of \eqn{eqn:density}. However, this results in an additional number of free parameters that produce an accurate fit, but with unrealistic model solutions.

The final spectrum in each pixel is then produced by integrating $E$ along the $y$ direction, resulting in a PPV spectral cube of size ($N_x$, $N_z$, $N_\varv$), where the last dimension corresponds to the observed velocity channels.

Our code includes a differentiable beam \reply{convolution} factor in the synthetic cube. However, this requires a higher spatial grid resolution to properly sample the dilution. This factor can vary depending on the quality of the observation, which has a significant impact on the required GPU memory and might necessitate a splitting strategy for the workload\reply{, including batching}, resulting in a substantial degradation of computational efficiency. Although \reply{in general} it would be possible to optimize \reply{the current} implementation, we found that for the specific case discussed in this work, the beam \reply{convolution} has a minimal impact on the final results; therefore, this improvement falls beyond the scope of the present paper.

\subsection{Free parameters of the model}
The steps described in \sect{sect:emission} require a series of free parameters. However, not all their parameters are treated independently, since we use for both the neutral and the ion the same versor matrix ($M$), the rotation angles ($\vartheta_{\rm y}$, $\vartheta_{\rm p}$, and $\vartheta_{\rm r}$), the ellipticity ($\delta$), and the collective object velocity  ($\varv_{\rm bulk}$). This is to mimic two consistent velocity fields (same direction locally, but different velocities), and two consistent density distributions (same rotation and ellipsoidal deformation, but different radial distributions). The free parameters are listed in \tab{tab:parameters} for reference, with the shared values indicated by an asterisk. 

All the model elements and operations mentioned so far are coded to leverage JAX vectorization and architecture, thereby reducing several computational bottlenecks. We also included only differentiable functions written to avoid vanishing gradients or unexpected non-differentiable configurations.

\begin{table}
    \caption{Model free parameters.}
    \centering
    \begin{tabular}{llcl}
        \hline
        Parameter& Description & Shared & Units\\
        \hline
        $M$ & Versor matrix coefficients & $*$ & -\\
        $\varv_{\rm bulk}$ & Object bulk velocity & $*$ & km\,s$^{-1}$\\
        $\varv_0$ & Gaussian $\varv(r)$ amplitude & & km\,s$^{-1}$\\
        $\sigma_\varv$ & Gaussian $\varv(r)$ width & & $L$\\
        $r_\varv$ & Gaussian $\varv(r)$ center & & $L$\\
        $a_\varv$ & Asymmetric pert. amplitude & & -\\
        $\varphi_0$ & Asymmetric pert. phase & & deg.\\
        $n_0$ & Gaussian $n(r^*)$ amplitude & & $L^{-3}$\\
        $\sigma_n$ & Gaussian $n(r^*)$ width & & $L$\\
        $r_n$ & Gaussian $n(r^*)$ center & & $L$\\
        $\delta$ & Ellipsoidal factor & $*$ & -\\
        $\vartheta_{\rm y}$, $\vartheta_{\rm p}$, $\vartheta_{\rm r}$ & Intrinsic rotation angles & $*$ & deg.\\
        $x_c$, $z_c$ & Offset in the plane of the sky && L\\
        $\sigma_{\rm t}$ & Line broadening & & km\,s$^{-1}$\\
        $\sigma_{\rm abs}$ & Absorption cross section  & & $L^2$\\
        \hline
    \end{tabular}
\tablefoot{Free model parameters. ``Shared'' column indicates parameters that are the same for \theneutral{} and \theion{}. Since the code is designed to operate in code units, we report only the parameter dimensions, where $L$ represents length, while ``-'' indicates dimensionless quantities. For the velocity and its related quantities, the code uses the same units as in the original observed PPV cubes, i.e., km\,s$^{-1}$.}\label{tab:parameters}
\end{table}

\section{Model optimization}\label{sect:fitting}
The model described in \sect{sect:model} can be reduced to a differentiable operator $\mathcal{M}(\theta):\mathds{R}^{N_{\theta}}\to\mathds{R}^{2\times N_x\times N_z\times N_\varv}$ that, given a set of $N_\theta$ parameters $\theta$, produces a collection of PPV cubes $C=\{C_0, C_1\}$, respectively for \theneutral{} ($C_0$) and \theion{} ($C_1$). Our corresponding target observations are PPV cubes with the same dimensions as the synthetic ones (analogously, $T=\{T_0, T_1\}$). We want to find the input parameters $\theta$ so that we minimize the loss
\beq\label{eqn:loss}
\mathcal{L} = ||\mathcal{M(\theta)}-T||_2 + \lambda_1\,P(\theta)\,,
\eeq
where the first term is a standard $L_2$-norm, and the second is a penalty factor scaled by a constant $\lambda_1$ to force the system into physically reasonable solutions. The penalty factor depends on the specific case, as it is only employed when the optimization easily falls into unphysical solutions. The default is $\lambda_1=0$. More details are reported in \sect{subsect:optimization}.

It is important to note that, unlike single-pixel and single-component models, our loss function encompasses, for both molecules, all the map pixels and velocity channels simultaneously, and the 3D model is optimized accordingly. In other words, the full two PPV cubes are simultaneously optimized.

To minimize the loss with respect to the input parameters, we utilize \textsc{Optax}, a framework to leverage the JAX architecture of our model and its GPU capabilities. Although the memory footprint is relatively large (approximately 60\,GB in production\footnote{All the tests discussed in this paper are performed with a NVIDIA A100 80\,GB card.}), the model evaluation is relatively fast (from 10 to 1000 models per second, depending on the number of velocity spectral channels and $y$-axes resolution). 

We choose arbitrary but reasonable initial parameter values and minimize the loss using an Adam optimizer \citep{Kingma2014} with a learning rate of $10^{-3}$. All the other optimizer's parameters are used as their default values. After a few thousand iterations, the loss becomes constant, and the optimization process no longer improves. Additionally, we visually inspect the spectral fitting to assess the quality of the minimization.

We observed that the choice of initial parameters in some cases influences the final solution, and we identified certain initial configurations that evolve into local minima, where the minimization process does not improve while resulting in a poor quality of the spectral fit. To avoid this issue and generalize the method, we have tested a few machine learning data augmentation techniques, which only provided a limited effect in avoiding such local minima  (see e.g., \citealt{Zhong2017,Xu2023,Kumar2024}).

In particular, we used random masking and random pixel masking. In the former, we randomly remove some PPV elements from the loss calculation, while in the latter, we completely remove all the channels of a specific set of pixels. These randomizations are performed every optimization epoch, and we tested different fractions of removed elements. This approach is conceptually similar to the classic deep neural network dropout.

Another data augmentation technique is to add noise to the modelled PPV. Although the synthetic spectral cube does not include any observational noise, it is possible to add to $\mathcal{M}$ a normally distributed noise with the same statistical properties as the observed one. This randomization is performed at every optimization epoch. However, as in the previous case, we noted that this has a limited impact on the generalizability of the optimization method.

Analogously, given the large number of parameters, we have no clear indication that our solution corresponds to a global minimum. As an additional test, we randomized the initial conditions to automatically find smaller losses; however, we found no general strategy to maximize the exploration of the initial conditions.

In general, we found that, given the large number of parameters, the initial guess plays a crucial role in the optimization process (see the example and the related discussion in \appx{sect:test}). In fact, in some configurations, certain parameters have the same effect on the final output, resulting in a solution degeneracy. This problem can be mitigated by the penalty factor in \eqn{eqn:loss}. Additionally, the initial guess may produce a non-convex configuration, in which case the optimizer is likely to fail or produce unphysical results. In theory, this can be verified by computing the sign of the eigenvalues of the Hessian of the model with respect to the parameters for each output. However, given the number of outputs, this is numerically intractable, at least with relatively standard techniques. \reply{Finally, to reduce the uncertainty related to the choice of the initial guess, different optimization methods could be used. To this aim, \textsc{Optax} allows us to switch between different optimization techniques, but our tests found that Adam was the most effective. However, we did not systematically explore the impact of varying the other optimizers' parameters. For this reason, hyperparameter optimization techniques might improve the optimization efficiency.}

\section{Results}\label{sect:results}
\subsection{Observed spectra}
\begin{figure*}
\centering
    \includegraphics[width=.6\textwidth]{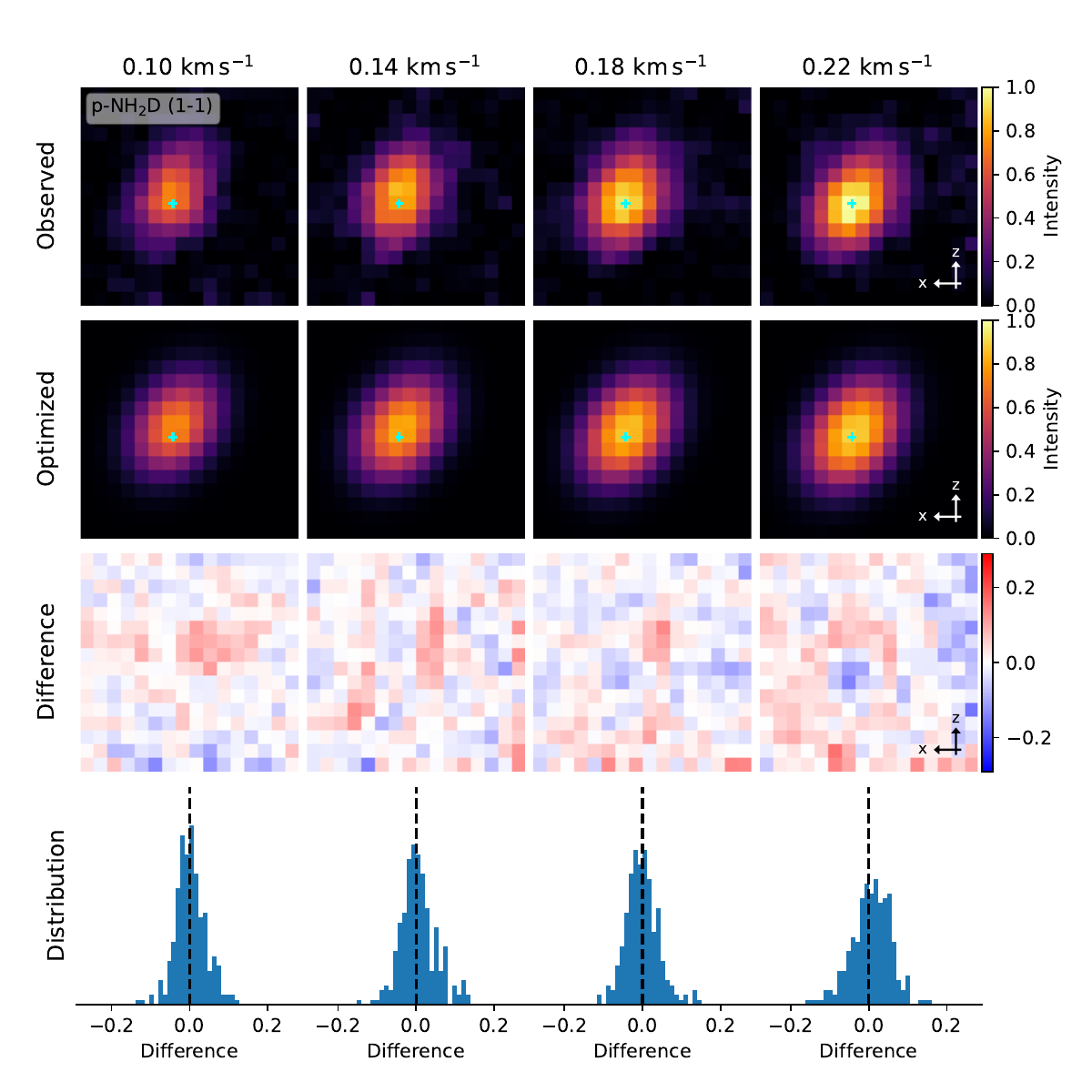}
    \caption{First and second rows are the observed and modelled intensity maps for some of the brightest selected velocity channels of \theneutral{}. The values are normalized to the same global maximum value of the PPV cube. The cyan cross is the position of the center as found by the optimizer, described by the $x_{\rm c}$ and $z_{\rm c}$ parameters (slightly different for each molecule, see \fig{fig:maps1}). The third row is the difference between the optimized and observed maps. The last row represents the density probability distribution of the difference maps.}
        \label{fig:maps0}
\end{figure*}

\begin{figure*}
\centering
    \includegraphics[width=0.6\textwidth]{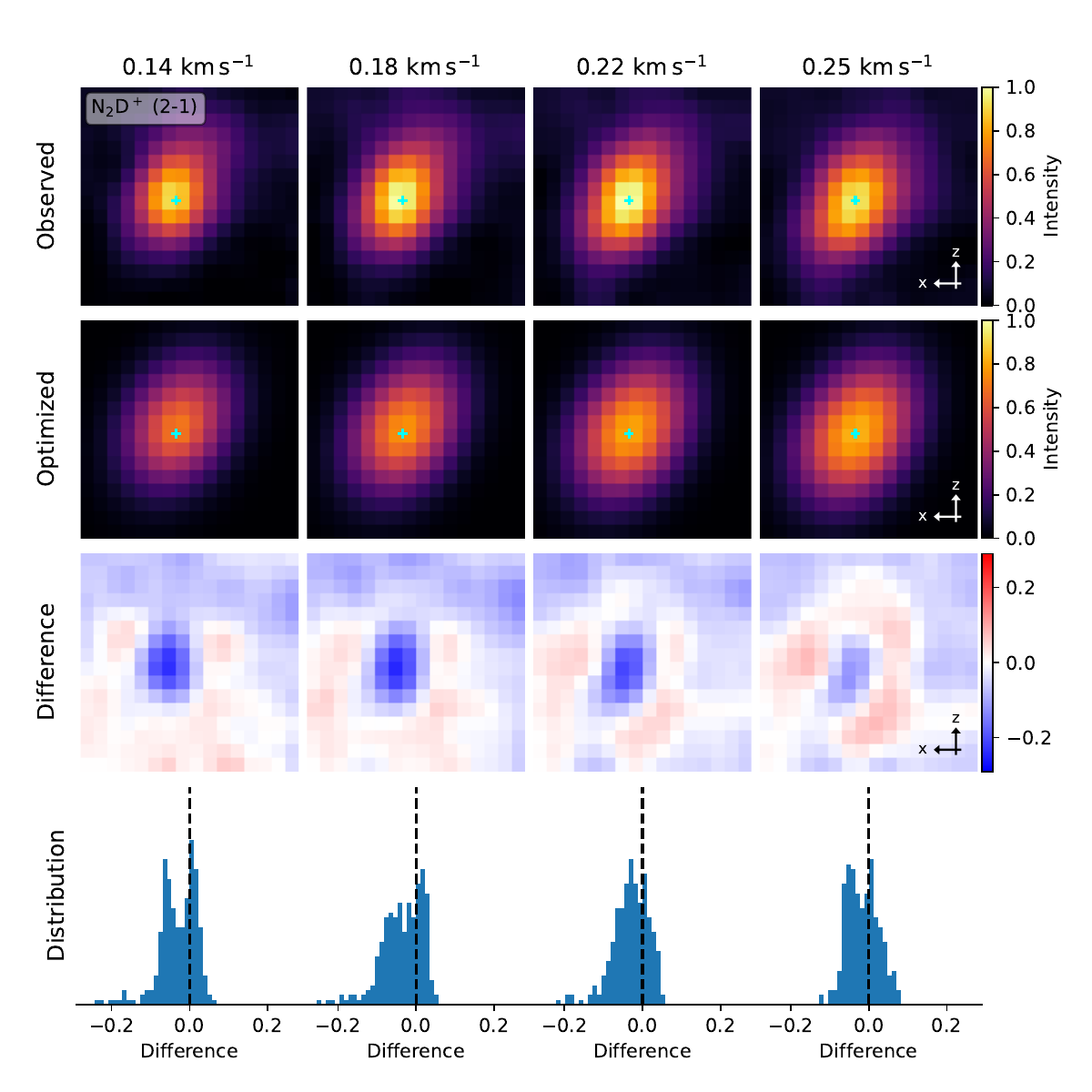}
    \caption{Same as \fig{fig:maps0} but for \theion{}.}
        \label{fig:maps1}
\end{figure*}

As an application example, we employ two observed PPV cubes of two spectral maps of the prestellar core L1544, specifically \theneutral{} (1-1) and \theion{} (2-1), from \citet{Arzoumanian2025}. An example is reported in the top row of \fig{fig:maps0} and \ref{fig:maps1}. They consist of two spatial axes in right ascension and declination, and a spectral axis expressed as LOS velocity. The pixel scale is $8$\,arcsec in both spatial directions and the reference velocity is \(7.2\,\mathrm{km\,s^{-1}}\). For \theneutral{} the spacing of the 1842 velocity channels is \(26.58\,\mathrm{m\,s^{-1}}\), the rest frequency of the reference observed transition is $110.153$\,GHz, and the beam is circular with $23.5$\,arcsec.
Instead, for \theion{}, the spacing of the 1817 velocity channels is $37.97$\,m\,s$^{-1}$, the rest frequency of the reference transition is $154.217$\,GHz, and the beam is $23.5$\,arcsec.
The data were observed at the IRAM-30m telescope (Pico Veleta, Spain). The spectral fit using \textsc{PySpecKit} in \citet{Arzoumanian2025} using 35 (\theneutral{}) and 26 (\theion{}) hyperfine lines, suggests the presence of a velocity shift between these two molecules. According to their analysis, the accuracy of the spectroscopical laboratory data (5-8\,m\,s$^{-1}$) when compared to the resolution of the observed data, excludes that the observed difference of the centroid is produced by the rest frequencies obtained. In this work, we use the original data of \citet{Arzoumanian2025}.

To minimize the memory impact, we preprocess the PPV cubes to match an arbitrary spectral grid of 256 velocity channels within the range -5 to 5\,km\,s$^{-1}$, i.e., with a resolution of 39.06\,m\,s$^{-1}$. To include all the relevant spectral features in the chosen range, and to reduce the number of relevant velocity channels, we subtract an arbitrary approximation of the object velocity of 7\,km\,s$^{-1}$ for both the molecules, and we linearly interpolate the original quantities on the desired grid. We are aware that the change in resolution and the interpolation strategy might affect the interpretation of the observations, especially when the relevant features (or their differences) are comparable to the grid scale. However, for the current application, we found that our interpolation has a limited impact on the final results.
The observed PPV cubes are normalized to their absolute maximum, and therefore, the solutions found will be in arbitrary code units, while the velocity units are unaffected (i.e., km\,s$^{-1}$).

\subsection{Optimization}\label{subsect:optimization}
As indicated in the previous section, the original analysis by \citet{Arzoumanian2025} of the data observed in L1544 reveals a velocity difference between the two molecular components. In particular, along the LOS, \theneutral{} is moving faster toward the center than \theion{}. A textbook symmetrical infall drift would result in a velocity difference sign inversion along the major axis of the projected prestellar core, since the closest and the farthest sides of the object present a different blue/red shift effect. For this reason, one possible solution to reproduce such observations is to have an actual sign inversion in the velocity differences in the model that compensates for the expected one. In other words, the fastest neutrals are in the closest region, and vice versa on the opposite side with respect to the center of the prestellar core. This is the preferred solution by the optimization, but we consider this solution to be unphysical, and therefore, we define a penalty factor in \eqn{eqn:loss} to constrain the magnitude of the velocity vector of \theneutral{} to be larger than or equal to the ion's analogous. In particular, we constraint the input parameters $\theta$ using $P(\theta)=\langle[\min(|\varv|_{\rm neutral}-|\varv|_{\rm ion}, 0)]^2\rangle$, where this is the average of the square of the negative differences of the magnitudes of the three-dimensional velocity structures for \theion{} and \theneutral{}, respectively\footnote{The $\min$ function is an element-to-element minimization equivalent to numpy's minimize.}. The scaling factor in \eqn{eqn:loss} after a few tests has been set to $\lambda=10$. In other words, this term deviates the loss from zero when the condition is not satisfied.
This constraint forces the optimizer to find a solution that produces the observed spectra, but with a global positive velocity difference between the neutral and ionized species.

Another constraint we have to impose to avoid unphysical solutions is to limit the density ellipsoid prolateness/oblateness in the range $\delta\in[0.5, 1.5]$ using a standard hyperbolic function\footnote{In particular, $\delta = 1 + 0.5\,\tanh\left[5(\delta' - 1)\right]$\,.}. Without this constraint, the optimizer tends to increase $\delta$ indefinitely, thereby flattening the ellipsoid to nullify the density of the off-plane regions and reduce the dimensionality of the problem.

To reduce the number of free parameters and the occurrence of unphysical velocity fields, we limit the number of non-zero elements in the versor matrix $M$. Instead of 9 independent elements, we use the same parameter for the three diagonal values ($M_{xx}=M_{yy}=M_{zz}$), representing a spherical infall component, and for the antisymmetric/orthogonal components around the $z_o$-axis (-$M_{xy}=M_{yx}$), representing a rotating field. All the other elements are set to zero. This will limit the geometrical configuration to have only two parameters that control the infall and the rotational component in the object coordinate system.

We provide the initial parameter guess values in \tab{tab:initial}. The optimization takes less than 10$^4$ epochs to minimize and reach a constant loss. It is worth noting that, although the minimized solutions fairly resemble the observed PPV cubes, with the current implementation and the given model, we cannot be certain whether this is a global minimum or a unique solution. 

\begin{table}
    \caption{Initial and final values of the parameters describing the model.}
    \centering
    \begin{tabular}{lrrr}
        \hline
        Par.& Init. & \multicolumn{2}{c}{Final}\\
        \hline
        & {\tiny Both} &{\tiny Neut.} & {\tiny Ion}\\
        \hline
        $M_{\rm in}$  &  0.1  & \multicolumn{2}{c}{-0.09}\\
        $M_{\rm rot}$  &  0.0  & \multicolumn{2}{c}{-0.04}\\
        $\varv_{\rm bulk}$  &  0.0  & \multicolumn{2}{c}{0.22}\\
        $\varv_0$  &  0.001 &  0.26 & 0.16\\
        $\sigma_\varv$  &  0.1 &  0.49 & 0.50\\
        $r_\varv$  &  0.5  &  0.91 & 0.90\\
        $a_\varv$  &  0.0  &  0.04 & -0.60\\
        $\varphi_0$  &  0°  &  67° & -76°\\
        $n_0$  &  1.0  &  0.59 & 0.71\\
        $\sigma_n$  &  0.03  &  0.38 & 0.56\\
                \hline
    \end{tabular}
    \vline
    \begin{tabular}{lrrr}
        \hline
        Par.& Init. & \multicolumn{2}{c}{Final}\\
        \hline
        & {\tiny Both} &{\tiny Neut.} & {\tiny Ion}\\
        \hline
        $r_n$  &  0.3  &  -0.21 & -0.56\\
        $\delta$  &  1.0  & \multicolumn{2}{c}{1.50}\\
        $\vartheta_{\rm y}$  &  -45°  & \multicolumn{2}{c}{-89°}\\
        $\vartheta_{\rm p}$  &  45°  & \multicolumn{2}{c}{31°}\\
        $\vartheta_{\rm r}$  &  -45°  & \multicolumn{2}{c}{-57°}\\
        $x_{\rm c}$  &  0.0  &  0.15 & 0.12\\
        $z_{\rm c}$  &  0.0  &  -0.07 & -0.04\\
        $\sigma_{\rm t}$  &  0.1  &  0.07 & 0.10\\
        $\sigma_{\rm abs}^{(a)}$  &  10.0  &  3.70 & 4.10\\
        {} & {} & {} &\\
        \hline
    \end{tabular}

\tablefoot{ $M_{\rm in}=M_{xx}=M_{yy}=M_{zz}$ is the infall component (negative values represent an infall), while $M_{rot}=-M_{xy} = M_{yx}$ is the rotational component. Initial values are the same for \theneutral{} (Neut.) and \theion{} (Ion), while final values can differ, see \tab{tab:parameters}. $(a)$ $\sigma_{\rm abs}$ is in units of $10^{-3}\,\Delta y= 2N_y^{-1}$, i.e., $L^{3}$ in code units.}\label{tab:initial}
\end{table}

\begin{figure}
\centering
    \includegraphics[width=0.48\textwidth]{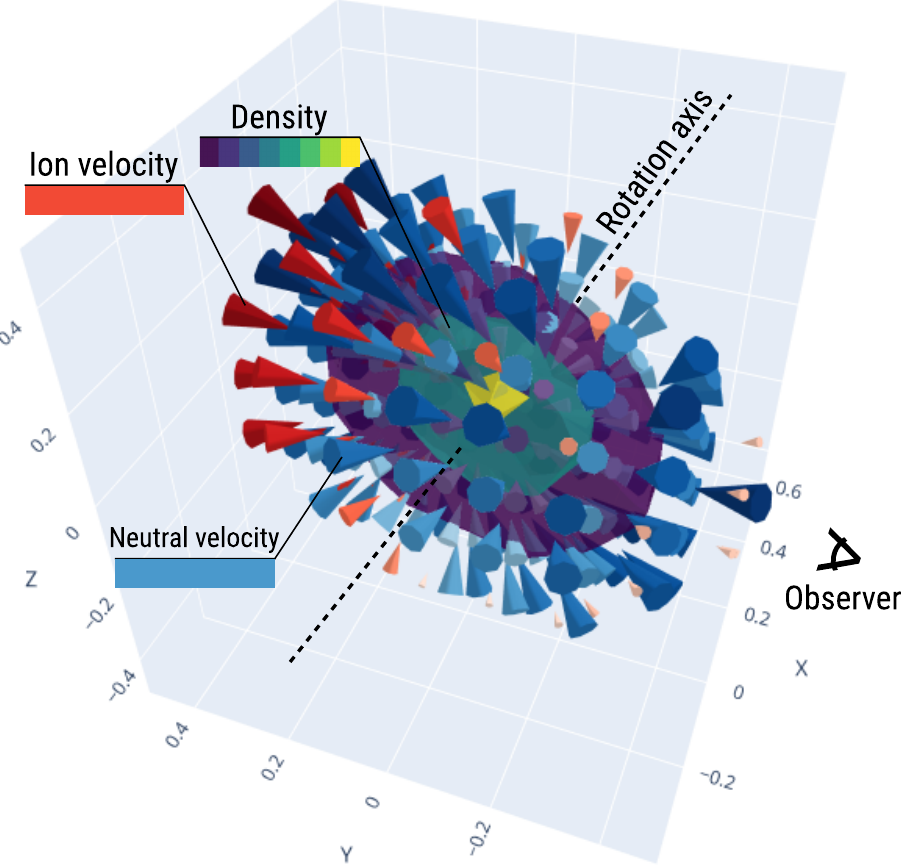}
    \caption{Three-dimensional schematic representation of the velocity vector fields (blue cones for \theneutral{} and red for \theion{}) and density isocontours (the colorscale in viridis is arbitrarily chosen to show the shape of the oblated spheroid). We include the position of the observer at $y=-1$ and the rotation axes, i.e., the yaw axis (rotation axis of $\vartheta_{\rm y}$). Note that the velocity difference is asymmetric, and the red cones closer to the observer overlap the blue ones. For the observer, the object appears as in Figs.\,\ref{fig:maps0} and \ref{fig:maps1}. Also, consider that the cones are reported to show the global properties of the vector field, but the observed velocity depends on the corresponding emitting density. To this aim, we report in \appx{sect:slices} the velocity and density slices at $x=0$ and $z=0$ (\fig{fig:slice_xz}), with the corresponding velocity difference (\fig{fig:slice_dv}).}
        \label{fig:3dplot}
\end{figure}

The optimized model found (see final values in \tab{tab:initial} and the schematics in \fig{fig:3dplot}) is an oblate ellipsoid rotated both in the plane of the sky and with respect to the position angle, similarly to what is expected from the observations of L1544 \citep{Ciolek2001,Rawlings2024}, where the velocity field is a mixture of infall and rotation, with infall dominating the overall motion (cf. \citealt{Tafalla1998}, \citealt{Ohashi1999}, and \citealt{Williams2006}). The velocity radial distributions of the two molecules are almost overlapping, but with a spatial asymmetry around the center of the object, as shown in \fig{fig:3dplot} by the different overlap between the blue and red vectors (compare the region closest to the observer with the farthest). This will be discussed in detail later. The density distributions peak almost in the center of the core (isocountours of \theneutral{} in \fig{fig:3dplot} and \fig{fig:slice_xz}), with \theion{} probing the innermost region (see \tab{tab:initial} and \fig{fig:ngas_hist}), which can be interpreted as a larger critical density (in fact, the expected value for \theion{} is $4\times10^5$\,cm$^{-3}$ compared to $1.4\times10^5$\,cm$^{-3}$ of \theneutral{}). Finally, the two molecules exhibit similar velocity dispersions (as also reported by \citealt{Arzoumanian2025}) and absorption, and a small position offset in the plane of the sky (cyan crosses in \fig{fig:maps0} and \ref{fig:maps1}).

To visualize the quality of the final results, we report in \fig{fig:maps0} and \ref{fig:maps1} a comparison in intensity maps for some of the brightest velocity channels. For the sake of clarity, we omit the other channels that present a similar agreement. In the upper row, we show the actual observed data, while in the second row, their optimized counterpart. To estimate the error, we include a third row representing a map of the absolute difference between the observed and modelled maps. The last row displays the probability density distribution of the error values from the previous row. We note that both molecules show an agreement between the observed and modelled maps. In the case of \theion{} (\fig{fig:maps1}), we select some of the channels that exhibit the largest difference in the central region, but this difference rapidly changes sign when changing velocity channels (see the pattern variation of the error in the third row). This is probably due to the intrinsic discrepancies between the model and the actual object, which likely has specific inhomogeneities and substructures that cannot be captured by our simplified model (cf. \citealt{Lin2022}).

\begin{figure}
\centering
    \includegraphics[width=0.48\textwidth]{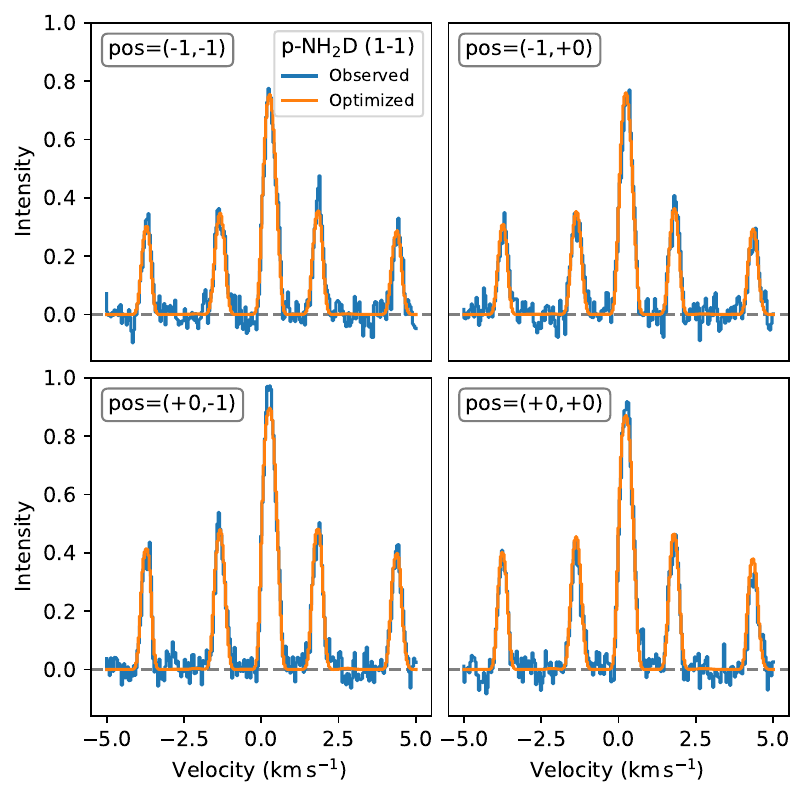}
    \caption{Comparison between observed (blue) and measured spectra (orange) of some selected pixels for \theneutral{}. The offset indicated in the ``pos'' box is in pixel units with respect to the central position, i.e., \mbox{(0, 0)} corresponds to (8, 8) within the 16$\times$16 pixel map. The intensity is the units of the observed PPV, normalized to their absolute maximum.}
        \label{fig:spectra0}
\end{figure}

\begin{figure}
\centering
    \includegraphics[width=0.48\textwidth]{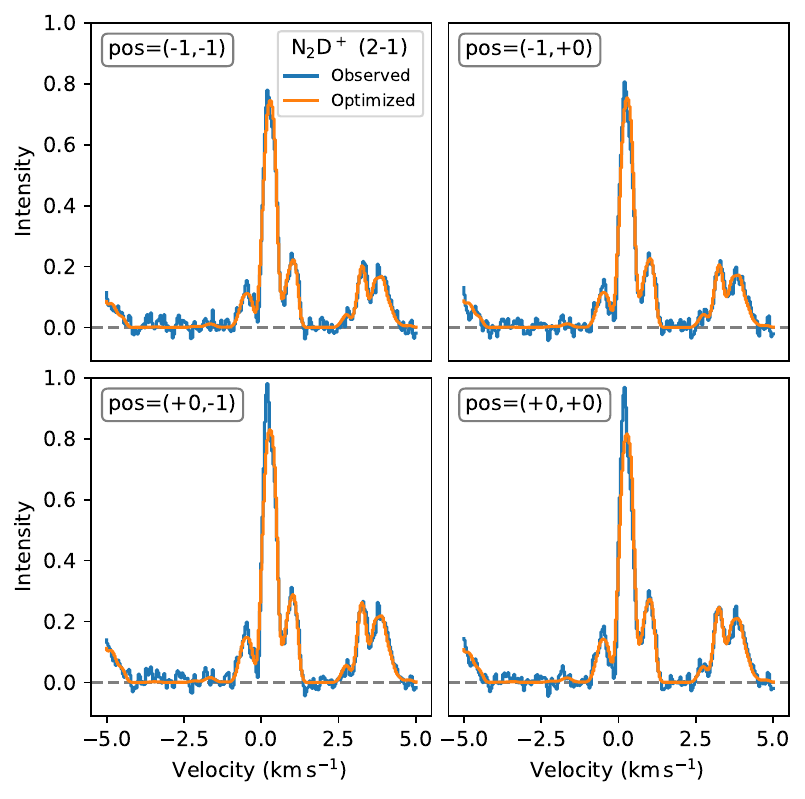}
    \caption{Same as \fig{fig:spectra0} but for \theion{}.}
        \label{fig:spectra1}
\end{figure}

For some of the central pixels, as shown in \fig{fig:spectra0} and \ref{fig:spectra1}, we also present a comparison between the observed spectrum (blue lines) and the one from the optimized model (orange lines). We note that most of the features are reproduced, especially for \theneutral{}, where some of the secondary features are different from the observed ones. In the case of \theion{}, the brightest peak is underestimated, depending on which pixel is considered. Similar considerations apply to the pixels not reported here, as can be inferred from \fig{fig:maps0} and \ref{fig:maps1}, where the error maps show some patterns, suggesting again that the intensity variations might be due to the model's lack of characteristic asymmetries, which may imply that it is incapable of reproducing a real object in all its nuances and details.

\begin{figure*}
\centering
    \includegraphics[width=0.75\textwidth]{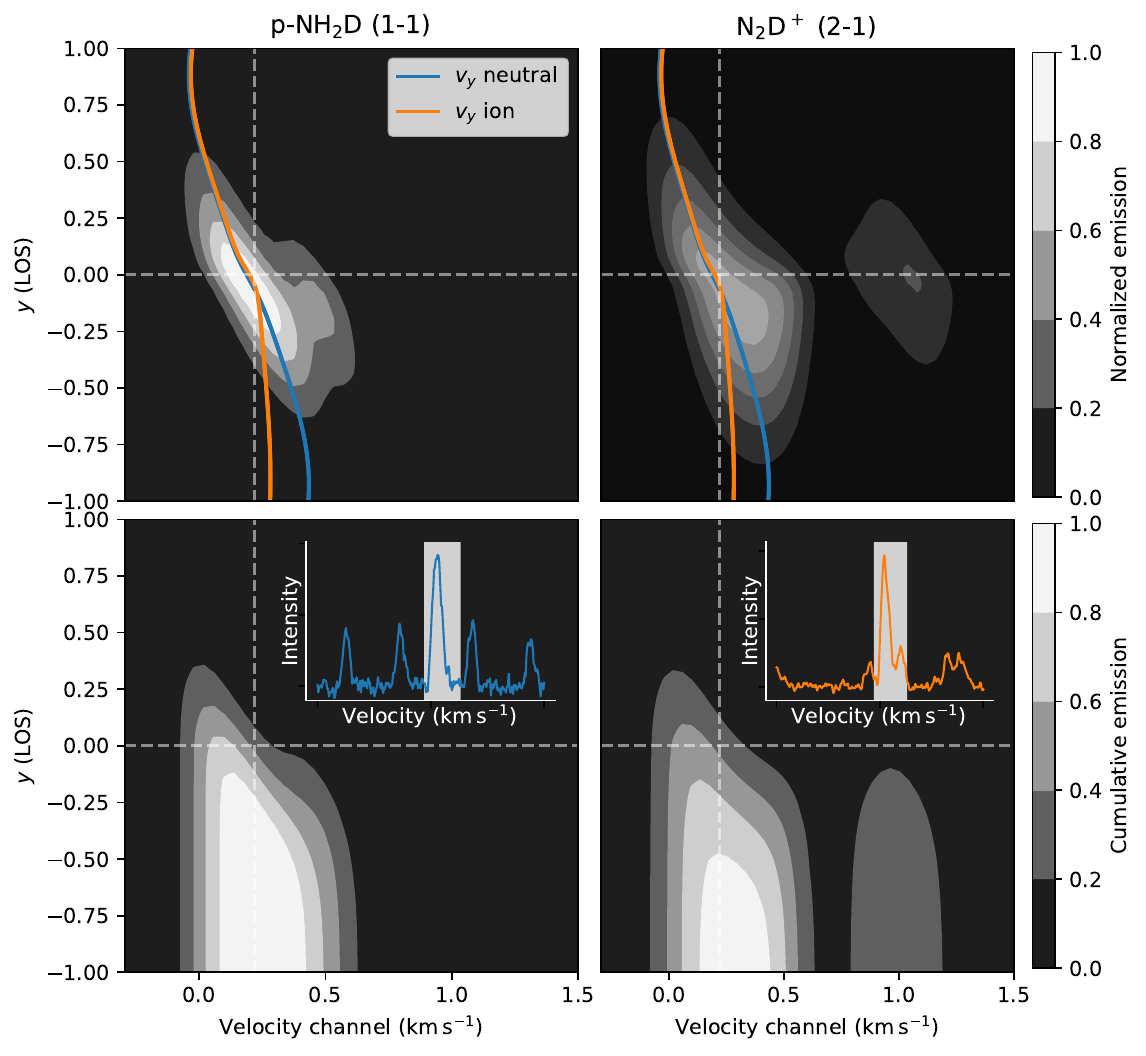}
    \caption{Top panels: effective emission intensity (i.e., including the absorption along the LOS) for \theneutral{} and \theion{}, at the position of the (8, 8) pixel within the 16$\times$16 pixel map. Values are normalized to the same maximum. The blue and orange lines represent the corresponding $y$ velocity components of \theneutral{} and \theion{}, respectively. The vertical dashed line is the bulk velocity of the object. Lower panels: cumulative effective emission along the LOS, where the observer is located at $y=-1$. All the panels are a zoomed part of the velocity channels (from -0.3 to 1.5\,km\,s$^{-1}$), while the $y$ coordinate covers all the model box. As a reference, we included the full spectrum in the insets, with the zoomed region highlighted.}
        \label{fig:emission}
\end{figure*}

As previously mentioned, a notable feature observed in the observed data of L1544 from \citet{Arzoumanian2025} is the absence of sign inversion in the velocity difference between ions and neutrals along the major axis, which is instead expected in the case of a textbook symmetric infall of two species with different velocities. In our case, we found that large-scale asymmetries in the velocity distributions might cause the observed effect, as modelled in \eqn{eqn:velocity}, represented by the different vector sizes between the two molecules in \fig{fig:3dplot}. The maximum, mean, and dispersion of the density-weighted velocity difference between these two components along the LOS ($\Delta \varv_y$) are around 0.18, 0.011, and 0.023\,km\,s$^{-1}$, respectively (see \fig{fig:slice_dv}). To quantify the difference, in \fig{fig:emission}, in the upper panels, we show the effective emission for the central pixel, i.e., considering the absorption along the LOS, as in \eqn{eqn:emission}. The contours indicate the emission in a given velocity channel (zoomed here in the range -0.3 to 1.3\,km\,s$^{-1}$) and at a given position $y$ along the LOS, with the observer located at $y=-1$ in code units. The blue and the orange lines are the $y$-component of the velocity of the emitting gas at a given position, respectively, for \theneutral{} and \theion{} (i.e., the LOS velocity component of the emitting grid element). The lower panels report the cumulative emission that contributes to the final spectra measured by the observer. In all the panels, the vertical dashed line indicates the bulk velocity. We note that for this specific solution, the velocity profiles are asymmetric with respect to the center, presenting a larger velocity drift on only one side of the object, with the ion showing a slower infall with respect to the neutral. The contours in the upper panel illustrate how each component contributes to the final spectrum (i.e., including the absorption), which is determined by blending the various spectral emission features into the final observation. The cumulative effective emission that produces the observed spectra is reported in the filled contour in the lower panels.

This result is consistent with several observations of L1544, which show asymmetries in the chemical structure in various tracers \citep{Spezzano2017}, e.g., CS \citep{Tafalla1998,Ohashi1999}, \ce{CH3OH} \citep{Bizzocchi2014,Punanova2018}, and \ce{CH3CCH} \citep{Giers2025}, as well as in \ce{N2H+} and \ce{HCO+} when modelled in hydrodynamical simulations with coupled chemical evolution \citep{Sipila2022}.

It is important to note that this specific solution, where the velocity asymmetry explains the observed data, is only one of the possible outcomes and is constrained by the given geometrical model. In fact, we find similar results by moving the asymmetry term in square brackets from \eqn{eqn:velocity} to \eqn{eqn:density}. This can be interpreted as a symmetric velocity field that maps asymmetric density regions. As it is, this produces two completely symmetric and separated ``lobes'' with the two molecular density distributions, one closer and one farther from the observer.  We consider this solution, with completely separated density regions, unphysical. This happens because $a_\varv$ in \eqn{eqn:velocity} reaches values of unity or more, and the phase $\varphi_0$ reaches for the two molecules a relative difference of around $\pi$. Therefore, this unphysical behavior can be limited by constraining $a_\varv$ to positive values smaller than, e.g.,\,0.3. In this way, we reach a solution that is again symmetric in velocity and only partially asymmetric in density, although it is not completely unphysical, unlike the previous one.

This other set of solutions with density asymmetry suggests that, at least within the constraints of our model, there are only two physically acceptable configurations, either with a velocity asymmetry and therefore a difference in velocity between the ion and the neutral components, or with no difference in velocity, but a difference in density (or a blend of these two solutions). In all other cases, the solutions exhibit unphysical behaviour incompatible with the observed properties of L1544.

Similarly, the optimization might find unphysical solutions when additional parameters are introduced. For example, adding a translation parameter along the LOS for both the centers of the density ellipsoids will cause the optimization to push the two ellipsoids apart, up to the limits of the model box. Since the vector field is not defined outside the box, this will produce two hemispheres with separated velocity components that mimic the observed drift. Clearly, this solution is unphysical, but it suggests that, although additional parameters might improve the quality of the fit, they could lead to completely unphysical solutions.

We do not explore other geometrical distributions, including, for example, variations from the ellipsoidal shape. However, we do not exclude the possibility that specific geometries, possibly supported by additional observational data, might result in better data interpretation, but this is beyond the aims of the present paper.

\subsection{Limitations}\label{sect:limitations}
This model presents a series of limitations that are introduced for physical or computational reasons. For example, to limit the number of parameters, we chose the density and velocity distributions to be Gaussian functions. This simplifies the modeling and can be directly interpreted as a probability, but it may differ significantly from the actual gas distributions. Another limitation is forcing the versor matrix ($M$) to be constant throughout the entire domain, which reduces the model's ability to detect asymmetries. However, using a more complicated representation could produce unrealistic minima or reduce the optimizer's capability. In general, the necessary simplified parameterization is a limit, since it cannot capture the actual features of the observed astrophysical object. This could, in principle, be overcome by using fully differentiable \reply{and self-consitent} physical models \reply{(as for example combining a differentiable hydrodynamical code to the radiative transfer part of our model insdead of using the geometrical parameterization)}, but their complexity, hardware requirements, and the lack of efficient differentiable numerical solvers (e.g., for evolving the chemistry) limit this approach (however, see the codes discussed in \sect{sect:introduction}).

Another relevant shortcoming in our pipeline is the simplified radiative transfer, which assumes that the emitting gas is identical to the absorbing one, apart from a scaling factor that depends on the absorption of each molecular component ($\sigma_{\rm abs}$). An improvement would be to compute the actual population of the molecular excited levels, such as in LOC \citep{Juvela1997,Juvela2020}, RADEX \citep{VanDerTak2007}, and LIME \citep{Brinch2010}, which would significantly increase the numerical complexity of the code and likely reduce its numerical efficiency. Finally, the computational impact in terms of GPU memory is quite relevant: one of the advantages of JAX is its capability to perform all calculations with efficient vectorization directly in memory, but, on the other hand, this reduces the amount of data that can be computed, which, in our case, translates to the size of the model grid. For example, with our current hardware (80\,GB GPU), a $64\times64$ spatial grid in the plane of the sky would require a parallelization strategy, which is beyond the current code architecture. \reply{Batching is another approach that can be considered to reduce the memory requirements, and it is relatively compatible with our current code logic}.

\section{Conclusions}\label{sect:conclusions}
We have developed a fully differentiable JAX pipeline to produce synthetic PPV data cubes from a geometrical 3D model controlled by a variety of parameters. The pipeline enables optimizing the parameters to match observed PPV data cubes. We applied our pipeline to observations of \theneutral{} and \theion{} in L1544 from \citet{Arzoumanian2025}.\\

\noindent The main conclusions are
\begin{itemize}
    \item The pipeline is numerically very efficient and allows for optimizing the model in minutes. The main limitation is the memory impact on the GPU.
    \item We find a relatively good agreement between observed and modelled PPV cubes when velocity or density asymmetries are introduced in the model.
    \item In particular, the observed difference in velocity between the two chemical species can be interpreted as an asymmetric velocity difference, larger in the region closer to the observer (with a maximum, mean, and peak differences of 0.18, 0.01, and 0.023\,km\,s$^{-1}$, respectively). This is in agreement with the existing knowledge of the L1544 structure.
    \item The model requires some constraints (i.e., a penalty in the loss) to avoid unphysical solutions, for example completely detached density structure for \theneutral{} and \theion{}, or velocity difference inversions along the LOS.
    \item Similarly, the optimization process can be relatively ``creative'', finding completely unphysical solutions that perfectly match the observed PPV data. This suggests these purely geometrical analytical models are limited by the choice of free parameters and the details of the geometrical assumptions.
    \item On the other hand, this suggests that modelling a specific prestellar core might require a detailed description of its structure, which single-component models cannot capture, especially in the case of high spatial or spectral resolution observations.
\end{itemize}

Our work demonstrates that differentiable algorithms serve as a key tool for interpreting observational data. However, to be more effective, they should include consistent physical information. To this aim, the next steps will be to implement consistent chemical evolution through the use of differentiable chemical solvers and, where possible, consistent dynamical evolution by using differentiable hydrodynamical solvers. This work presents the core concepts of the methods, which can then be applied to a larger data set to improve our understanding of the 3D physical and chemical structure of dense cores.

\begin{acknowledgements}
TG, JEP, SS, SJ, and PC gratefully acknowledge the support of the Max Planck Society. This work was
supported by the NINS-DAAD International Personal Exchange Program (2023-2025). This program is a joint funding initiative of the National Institutes of Natural Sciences (NINS) in Japan and the German Academic Exchange Service (DAAD).
\end{acknowledgements}

%
%
\bibliographystyle{aa}
\bibliography{mybib}

\begin{appendix}
\section{Infall and rotational model example}\label{sect:test}
To describe some of the characteristics of our code, we introduce a relatively simple case reported in the public repository\footnote{The corresponding notebook is \texttt{test.ipynb} at \url{https://github.com/tgrassi/ppv_jax_public} commit \texttt{79b7450}.}.

\subsection{Model description and optimization}
\begin{figure*}
\centering
    \includegraphics[width=0.9\textwidth]{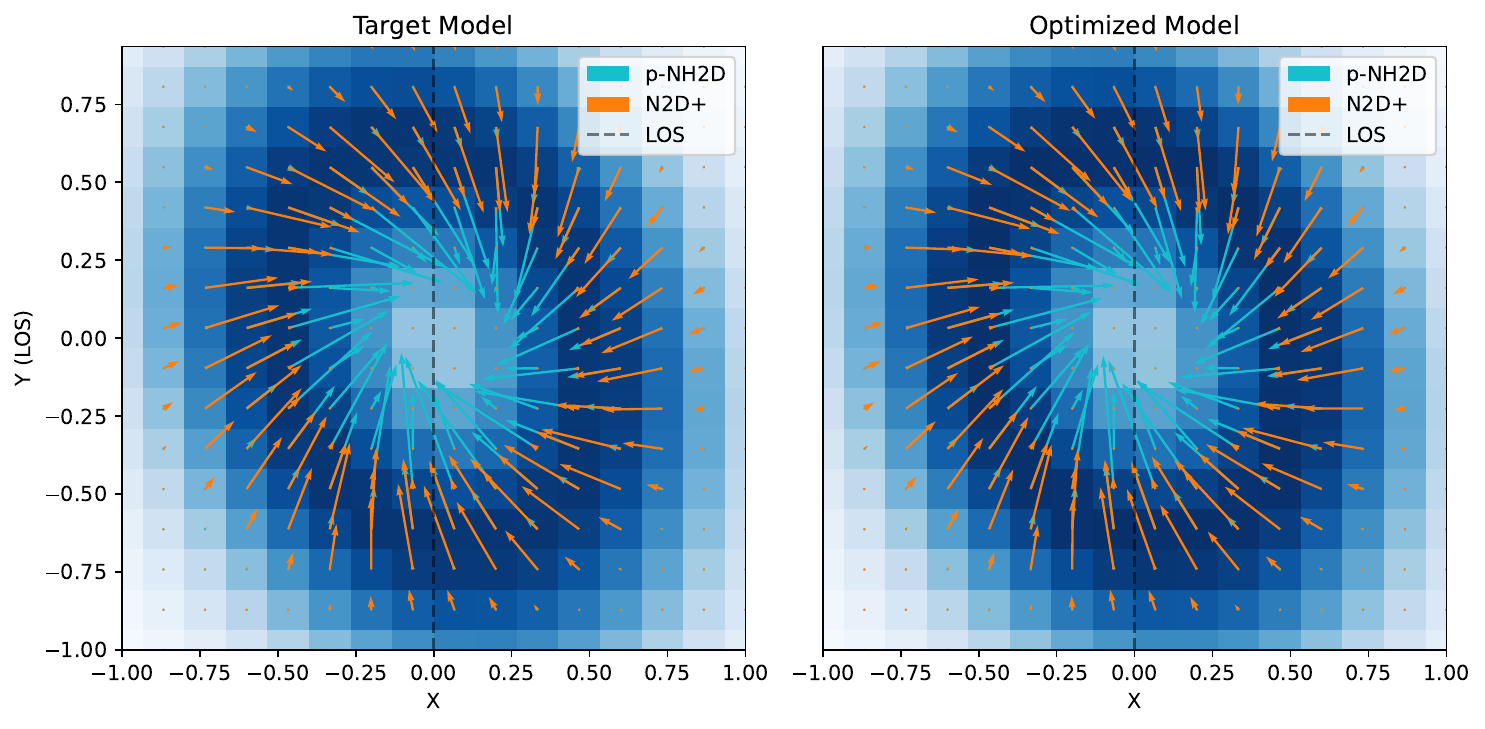}
    \caption{Slice at $z=0$ including the velocity vector field for \theneutral{} (cyan) and \theion{} (orange), and the gas density distribution of \theneutral{} (the blue color scale is in linear units normalized to the absolute maximum of both models). The left panel displays the target model, while the right panel shows the optimized model. The LOS is along the $ y$-axis, with the observer located at $y=-1$. The dashed line is the LOS in the central pixel (i.e., $x=z=0$).}
        \label{fig:xyslice}
\end{figure*}

This model features an equal combination of infall and rotation, with the emitting density exhibiting an oblate structure that includes an inner cavity (to mimic some inner region depletion) and some rotation on the three axes, albeit slightly different for the two tracers (\theneutral{} and \theion{}). The velocities are similar, but the Gaussians have different centers. There is a small bulk velocity, and we consider only 128 channels in the range -2 to 2\,km\,s$^{-1}$. We use the model routine to generate two target PPV cubes from these parameters. We then define an initial guess for the parameters and optimize the loss without the extra penalty factor to simultaneously reproduce the target PPVs. We interrupt the process when the loss is smaller than $10^{-8}$: the target spectra are perfectly reproduced, as well as the key feature of the structure, as shown in the comparison of \fig{fig:xyslice}, where the density of \theneutral{} and the velocity fields of both molecules are reported in the $z=0$ plane.

\subsection{Final parameters analysis}
\begin{figure*}
\centering
    \includegraphics[width=0.95\textwidth]{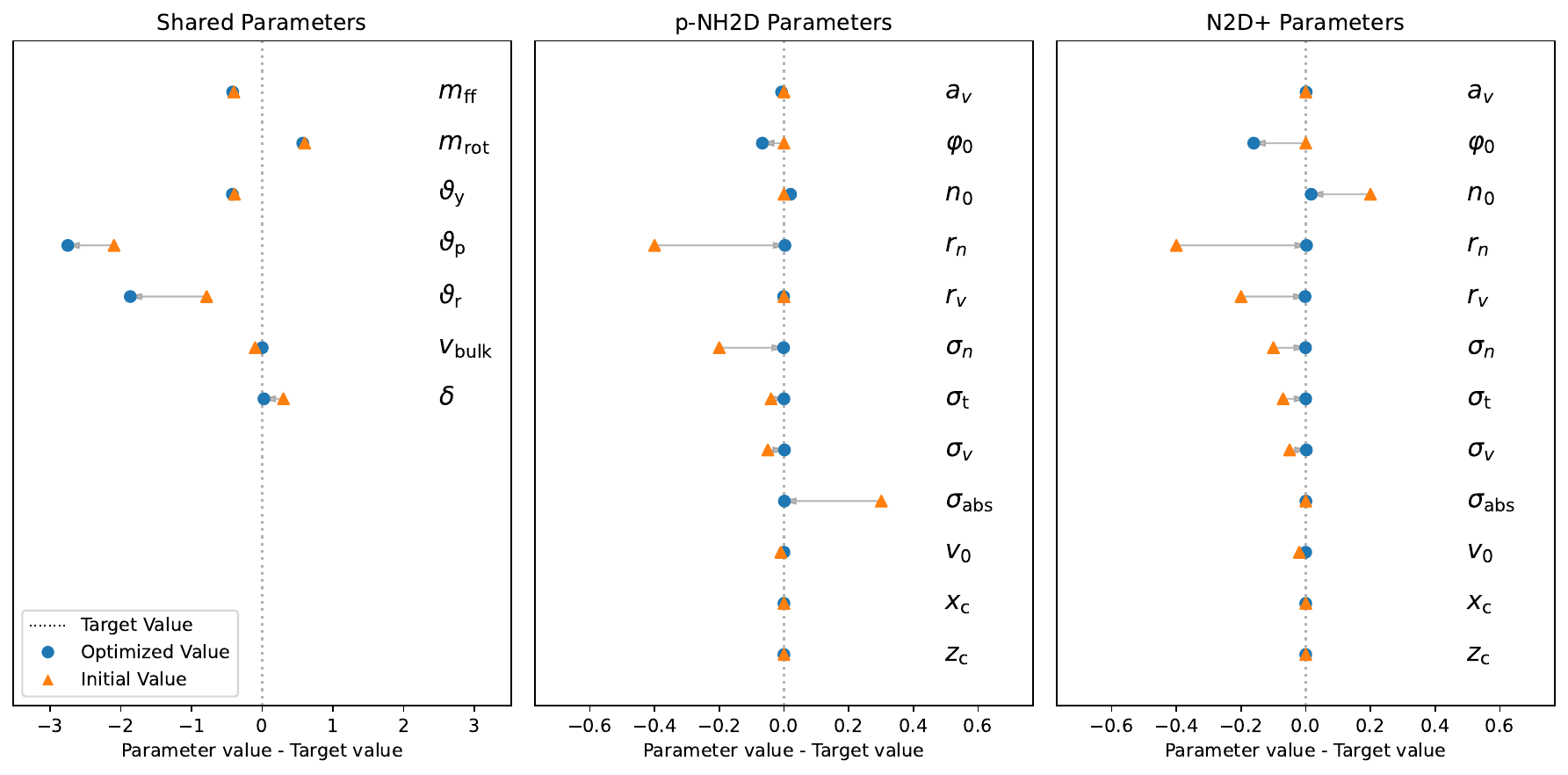}
    \caption{Difference between the initial (orange triangles) and final values (blue circles) with the target value (vertical line), i.e., zero (vertical line) is the perfect match with the expected value. In the first left, the parameters shared with both molecules are reported, while the centra and right panels report the parameters of \theneutral{} and \theion{}, respectively. The size of the various changes is not comparable, as different values have varying impacts on the final quantities (see \fig{fig:derivatives}).}
        \label{fig:parameters}
\end{figure*}

We also noted that, despite the PPV cubes being perfectly reproduced, some parameters are optimized with values different from the target ones, as shown in \fig{fig:parameters}, where we compare the differences between the target values and the optimized and initial values. While most of the parameters are correctly reproduced (blue circles on top of the dashed line), some parameters have different values (e.g., the rotation angles $\vartheta$), or they have not been modified at all (e.g., $m_{\rm ff}$ and $m_{\rm rot}$). This is related to the fact that not all parameters affect the final result, and that some parameters exhibit degeneracies, i.e., to obtain the optimized PPVs (in some specific configurations), the optimizer obtains the same effect by modifying only one or two or more parameters\footnote{As an example, in $a+b=1$,  to obtain the result, $a$ can be modified without changing $b$ or vice versa.}. A notable example is $m_{\rm rot}$, which the optimization found with a different sign with respect to the target value; i.e., the rotation direction is expected to be counterclockwise, contrary to what is in the target model. However, the optimizer using the angles $\vartheta$ flipped the object to match the expected direction. Analogously, affecting the other parameters, such as density and velocity distribution, the optimization does not need to change the actual values of $m_{\rm ff}$ and $m_{\rm rot}$, but only uses their signs\footnote{However, only for $m_{\rm ff}$, given the flip via the angles $\vartheta$ described before}. Other parameters, such as the phase of the asymmetry $\varphi_0$, which appears relatively large, are utilized by the optimizer to fine-tune the model. In fact, in this case, the amplitude of the asymmetry $a_\varv$ is negligible (the actual target value is zero), and therefore it plays a very minor role.

This parameter analysis suggests again that, although the final PPVs and the model structure are well reproduced, the parameters can be degenerate.

\subsection{Derivatives of the spectra with respect to the parameters}
\begin{figure*}
\centering
    \includegraphics[width=0.9\textwidth]{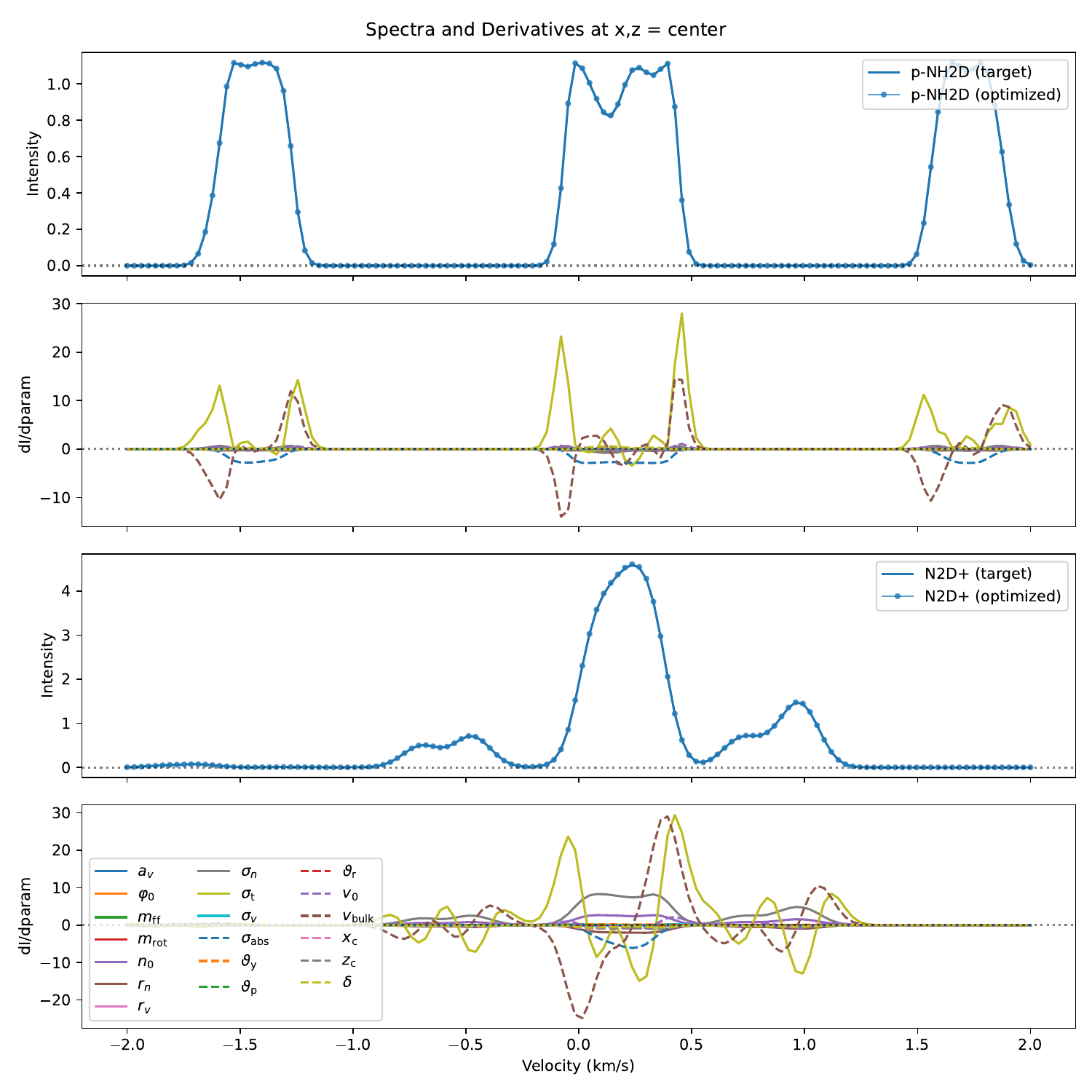}
    \caption{First and third panels: target (solid) and optimized emission spectra (solid with circle marker) in the central pixel $x=z=0$, for \theneutral{} and \theion{}, respectively. The lines overlap. Second and fourth panels: derivatives of the emission with respect to the given parameter $\theta_i$ reported in the legend, i.e., $\partial I(\varv)/\partial \theta_i$. Note that the absorption coefficient is actually $\log_{10}(\sigma_{\rm abs}\Delta y)$.}
        \label{fig:derivatives}
\end{figure*}

Not all the parameters contribute equally to the optimized PPVs. In fact, to inspect the role of the different parameters, we compute the derivatives of each output with respect to each parameter (i.e., the Jacobian), as reported in \fig{fig:derivatives}. Using JAX, computing the derivatives is numerically straightforward\footnote{To avoid materializing the full Jacobian in memory and thereby reduce its size, instead of \texttt{jax.jacobian}, we use the Jacobian-vector product \texttt{jax.jvp} with a zero vector with one only in the position of the given parameter $\theta_i$.}. The figure shows for \theneutral{} (first two panels from the top) and \theion{} (last two panels), the spectra of the central pixel $x=z=0$ (first and third panels) and the correspondent derivatives with respect to each parameter (second and fourth panels), i.e., $\partial I(\varv)/\partial \theta_i$, where $\theta_i$ is a given parameter. We found that, in this particular case at the final optimization step, the most important parameters are, the microturbulence $\sigma_t$, the bulk velocity $\varv_{\rm bulk}$, the absorption factor $\sigma_{\rm abs}$, here as the derivative of $\log_{10}(\sigma_{\rm abs}\Delta y)$, and the scaling and the width of the density gas distribution Gaussian function, respectively $n_0$ and $\sigma_n$. From \fig{fig:derivatives}, we note that the density parameters ($n_0$ and $\sigma_n$) and the absorption ($\sigma_{\rm abs}$) control the intensity of the convolved lines, because the first two parameters control the gas emission, while the other, by definition, the absorption (in fact, the derivative of the intensity is positive for the first two and negative for the latter). Instead, $\sigma_t$ controls the ``sides'' of the emission features, since for larger turbulence we expect to have broader lines. Finally, the bulk velocity $\varv_{\rm bulk}$ rigidly translates the features along the velocity axes, which explains why the derivative changes sign when comparing the left and the right ``sides'' of the emission features.

\subsection{Contribution to the observed emission}
\begin{figure*}
\centering
    \includegraphics[width=0.9\textwidth]{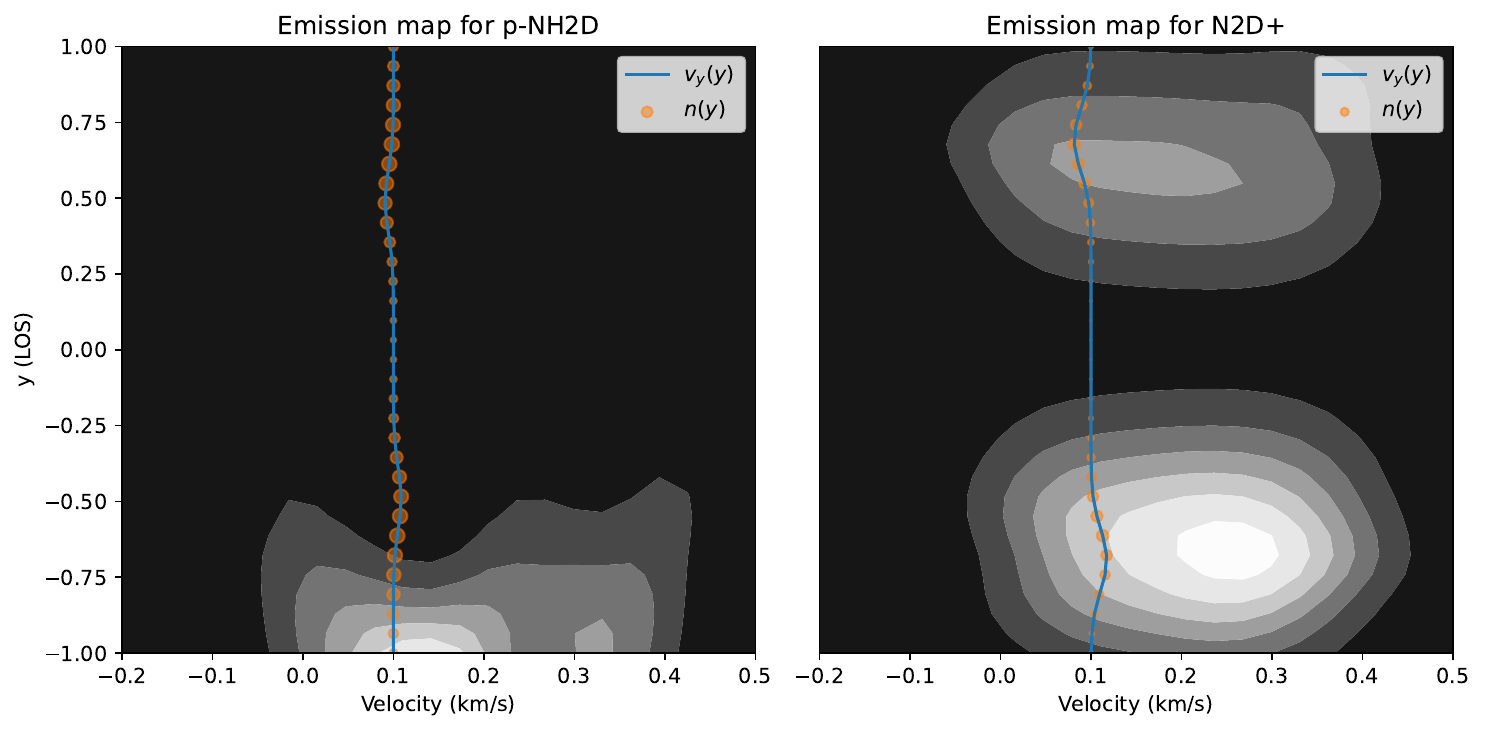}
    \caption{The filled gray color contours are the \theneutral{} (left) and \theion{} (right) effective emission (i.e., considering the absorption and normalized to the global maximum emission) for the grid elements along the LOS ($y$-coordinate) in the central pixel ($x=y=0$). The emission is plotted for each individual velocity channel in the -0.2 to 0.5\,km\,s$^{-1}$ range (horizontal axis). The solid blue line is the $\varv_y$ component of the gas velocity as a function of the LOS coordinate, while the area of the orange circles is scaled as the molecule density along the LOS. The observer is located at $y=-1$.}
        \label{fig:emission_test}
\end{figure*}

The code can store the effective emission (i.e., considering the absorption) for each element of the 3D grid and for both molecules. In \fig{fig:emission_test}, we report the final emission map (filled contours normalized to the global emission) along the LOS of the central pixel (i.e., $x=z=0$) and for different velocity channels in the -0.2 to 0.5\,km\,s$^{-1}$ range. The observer is located at $y=-1$. The solid line represents $\varv_y$ along the LOS, while the area of the orange circles is scaled as the corresponding density. We note that although the emission of \theneutral{} should scale as the density, given its relatively large absorption $\sigma_{\rm abs}$, the contribution comes only from the region closer to the observer. Conversely, \theion{} emission comes from regions at $y>0$, because it has a smaller absorption.\\

More plots can be found at \url{https://github.com/tgrassi/ppv_jax_public/test.ipynb}.

\section{Velocity and density projections}\label{sect:slices}
\begin{figure*}
\centering
    \includegraphics[width=0.48\textwidth]{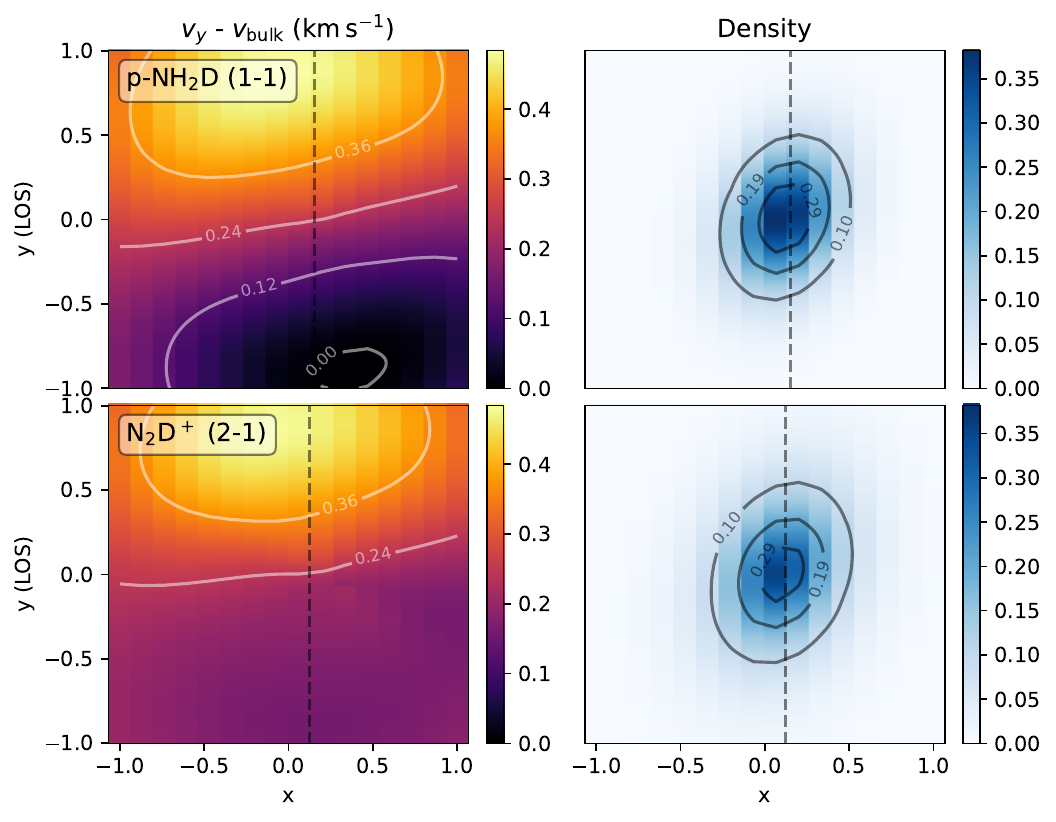}
    \includegraphics[width=0.48\textwidth]{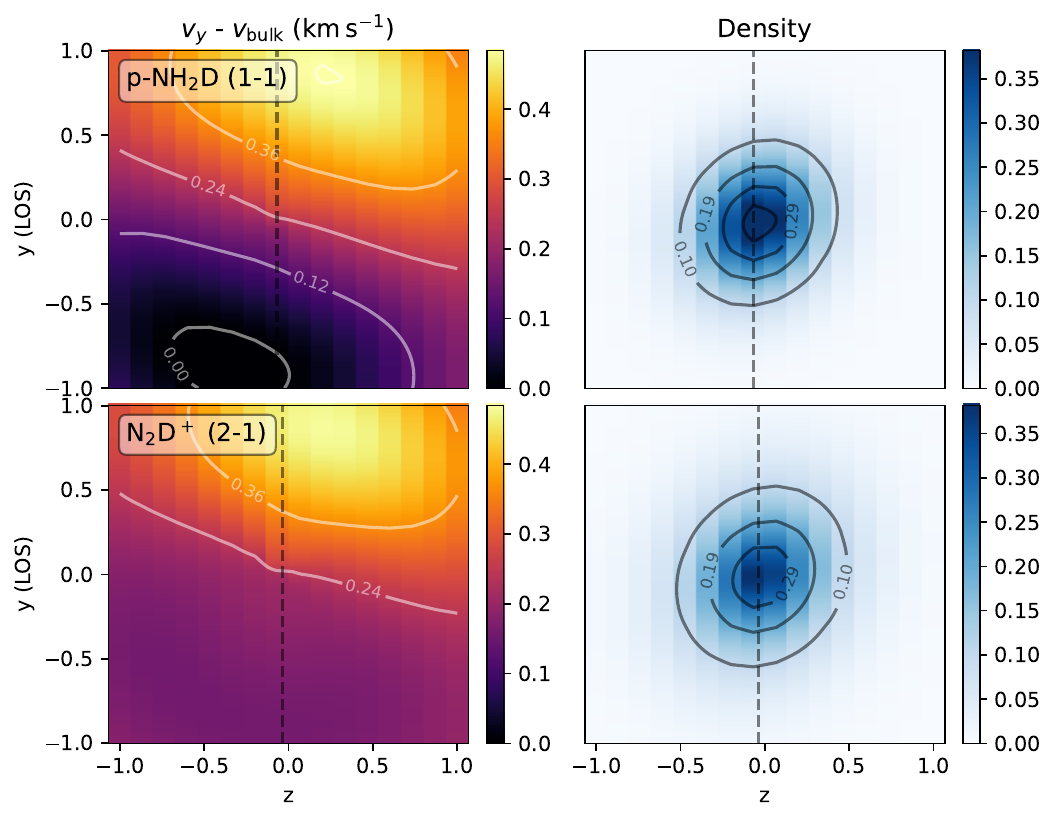}
    \caption{The left group of four panels shows a $z=0$ cut of the model ($x-y$ plane) reporting with a dark violet to yellow color scale the LOS component of the velocity (i.e., along the $y$ axis, $\varv_y$) for \theneutral{} (top rows) and \theion{} (bottom rows), with the corresponding density (shades of blue color scale). The right groups of panels report the same but for $x=0$ ($z-y$ plane). The dashed vertical lines represent the position of the center offsets, i.e., the $x_c$ and $z_c$ parameters. }
        \label{fig:slice_xz}
\end{figure*}

In \fig{fig:slice_xz} we report the magnitude of the component along the LOS ($\varv_y$) relative to the bulk velocity for \theneutral{} and \theion{} and the corresponding density in code units. The group of plots on the left is a cut of the model box at $z=0$ ($x-y$ plane), while the group on the right is a cut at $x=0$ ($z-y$ plane). The dashed lines represent the position of the center offsets, i.e., the $x_c$ and $z_c$ parameters. See \fig{fig:3dplot} for comparison.

\begin{figure*}
\centering
    \includegraphics[width=0.68\textwidth]{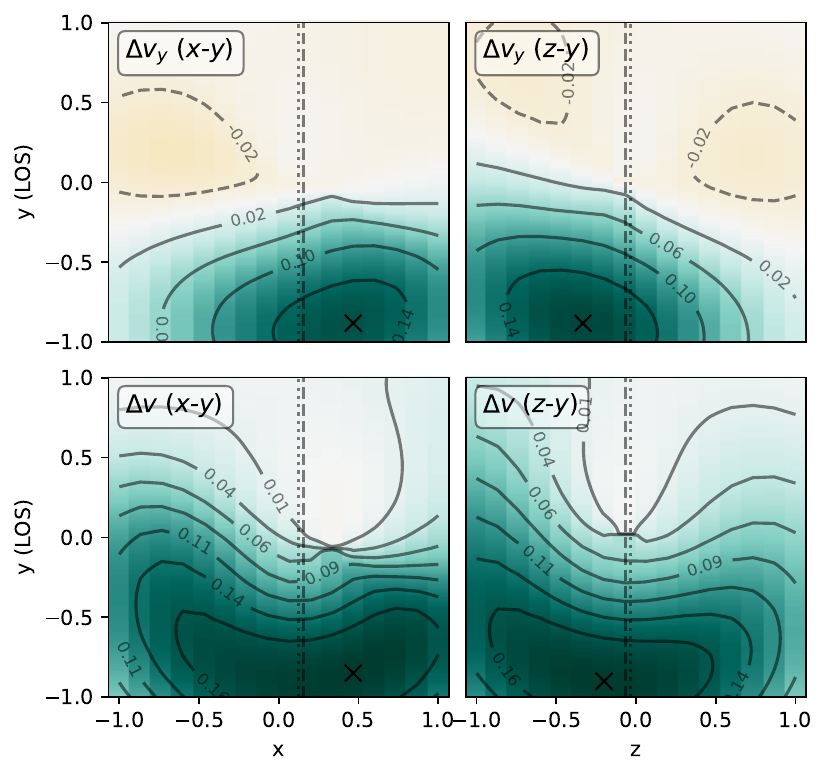}
    \caption{Velocity difference along the LOS, i.e., $\Delta \varv_y$ (upper panels) and absolute velocity difference, i.e., $\Delta \varv$ (lower panels). Left panels are slices for $z=0$, while right panels $x=0$. The color scale corresponds to the contour values. The vertical lines indicate $x_c$ (left panels) and $z_c$ (right panels) for \theneutral{} (dashed) and \theion{} (dotted). The peak of the color scale is 0.185\,km\,s$^{-1}$, the black crosses indicate the position of the maximum difference.}
        \label{fig:slice_dv}
\end{figure*}

In \fig{fig:slice_dv} we report the difference in the magnitudes of the velocity along the LOS ($\Delta\varv$, top panels) and for the velocity vector magnitude ($\Delta \varv$, bottom panels). The slices are $z=0$ ($x-y$ plane, left panels) and $x=0$ ($z-y$ plane, right panels).

\begin{figure*}
\centering
    \includegraphics[width=0.48\textwidth]{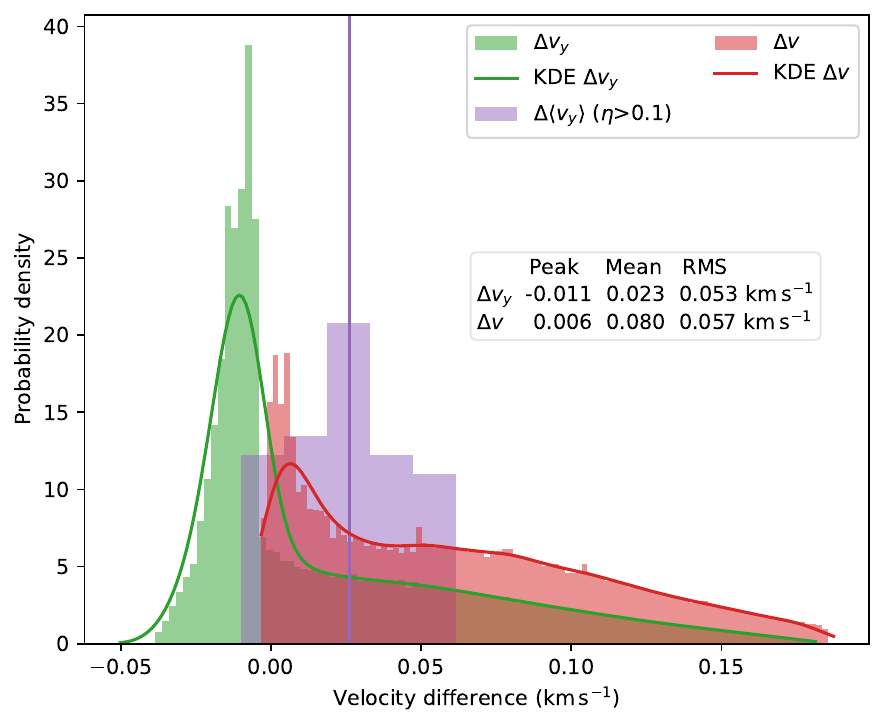}
    \includegraphics[width=0.48\textwidth]{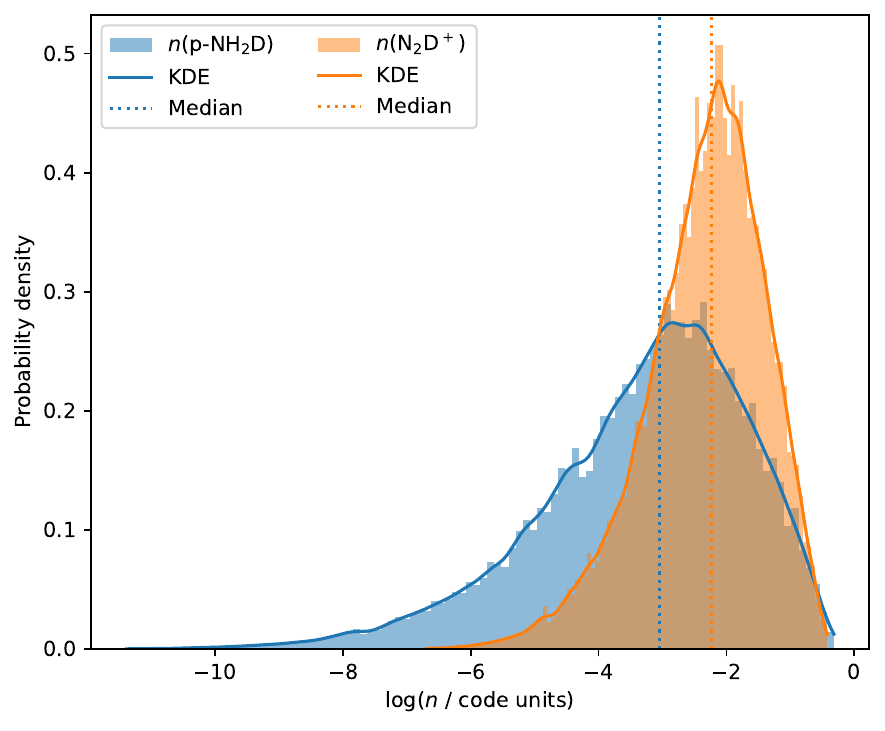}
    \caption{Left: probability density (histogram) and corresponding Gaussian KDE (using Scott's rule, solid lines) velocity difference distribution for the LOS component $\varv_y$ (green) and the velocity magnitude $\varv$ (red). All values are weighted by their corresponding number densities (i.e., regions that emit more have a larger weight in the distribution). In addition, we report (violet histogram) the difference in the density-weighted average $\varv_y$ along the LOS, excluding pixels where the normalized integrated column density ($\eta$) is less than 0.1 times its maximum. The vertical line is the mean of the distribution (0.029\,km\,s$^{-1}$). In the textbox, we also report the peak positions, the means, and the standard deviations of the two distributions. Right: probability density (histogram) and corresponding Gaussian KDE (using Scott's rule, solid lines) of the logarithm of the gas density of \theneutral{} (blue) and \theion{} (orange). The vertical dotted lines are the corresponding median values.}
        \label{fig:ngas_hist}
\end{figure*}

In \fig{fig:ngas_hist}, left panel, we measure the statistical properties of the velocity difference. In particular, we show the gas-density-weighted probability density for both $\varv_y$ (green) and $\varv$ (red). Note that $\varv$, by definition, has no sign, while its difference does, which explains the separation between the peaks of $\Delta \varv_y$ and $\Delta \varv$. The peak of the distributions, the mean, and the dispersion are $0.011$, $0.023$, and $0.053$\,km\,s$^{-1}$ for $\Delta v_y$, and $0.007$, $0.080$, $0.057$\,km\,s$^{-1}$ for $\Delta v$. The solid lines are the Gaussian Kernel Density Estimation (KDE), using Scott's rule. In the same figure we report the difference between $\langle \varv_y \rangle = \int_{-1}^1 n(y)\, \varv_y(y)\,\dd y\, / \int_{-1}^1 n(y)\,\dd y$ for both molecules (i.e., $\Delta \varv_y$), but only including regions where the normalized integrated column density $\eta = \int_{-1}^1 n(y)\,\dd y / \max\limits_{\forall\rm pixels}\left[\int_{-1}^1 n(y)\,\dd y\right]$ is larger than 10\%.

In \fig{fig:ngas_hist}, right panel, we report the probability density distribution of the logarithm of the gas density for \theneutral{} (blue) and \theion{} (orange). The dotted vertical lines indicate the median of the distributions. Also in this panel, the solid lines indicate the Gaussian Kernel Density Estimation (KDE), using Scott's rule.

\end{appendix}

\end{document}